\documentclass[twocolumn,fleqn,usenatbib]{mnras}

\usepackage{amssymb}	
 
\usepackage{newtxtext,newtxmath}

\usepackage[T1]{fontenc}

\DeclareRobustCommand{\VAN}[3]{#2}
\let\VANthebibliography\thebibliography
\def\thebibliography{\DeclareRobustCommand{\VAN}[3]{##3}\VANthebibliography}

\usepackage{graphicx}	
\usepackage{mathtools} 

\usepackage[noabbrev, nameinlink]{cleveref} 

\usepackage{bm}

\usepackage{csquotes}
\usepackage{savesym}
\usepackage{microtype}

\savesymbol{tablenum} 
\usepackage{siunitx}
\restoresymbol{SI}{tablenum}

\usepackage{threeparttable}

\usepackage{textgreek}

\usepackage[hashEnumerators,smartEllipses]{markdown}

\usepackage{caption}

\graphicspath{{./}{figures/}{figures-png/}}  

\providecommand{\abs}[1]{\left\lvert#1\right\rvert}
\providecommand{\norm}[1]{\left\lVert#1\right\rVert}

\newcommand{\patch}[1]{\SI{#1}{\degree}$\times$ \SI{#1}{\degree}}
\newcommand{\smallpatch}[1]{\SI{#1}{\arcmin} $\times$ \SI{#1}{\arcmin}}

\newcommand{\klos}{\text{$k_{\parallel}$}}
\newcommand{\kperp}{\text{$k_{\bot}$}}

\DeclareSIUnit\arcsec{arcsec}
\DeclareSIUnit\persqarcsec{$arcsec^{-2}$}
\DeclareSIUnit\arcmin{arcmin}
\DeclareSIUnit\cMpc{cMpc}  
\DeclareSIUnit\cGpc{cGpc}  
\DeclareSIUnit\deg{deg}
\DeclareSIUnit\sqdeg{$deg^2$}
\DeclareSIUnit\persqdeg{$deg^{-2}$}
\DeclareSIUnit\erg{erg}
\DeclareSIUnit\gauss{G}
\DeclareSIUnit\hour{h}
\DeclareSIUnit\hubble{\text{$h$}}
\DeclareSIUnit\jansky{Jy}
\DeclareSIUnit\lightyear{light-year}
\DeclareSIUnit\parsec{pc}
\DeclareSIUnit\rayleigh{Rayleigh}
\DeclareSIUnit\solarmass{\text{M$_{\odot}$}}
\DeclareSIUnit\Msun{\solarmass}
\DeclareSIUnit\year{yr}
\DeclareSIUnit\keV{\kilo\electronvolt}
\DeclareSIUnit\kpc{\kilo\parsec}
\DeclareSIUnit\mJy{\milli\jansky}
\DeclareSIUnit\uJy{\micro\jansky}
\DeclareSIUnit\mK{\milli\kelvin}
\DeclareSIUnit\uG{\micro\gauss}
\DeclareSIUnit\Gyr{\giga\year}
\DeclareSIUnit\MHz{\mega\hertz}
\DeclareSIUnit\Mpc{\mega\parsec}
\DeclareSIUnit\Gpc{\giga\parsec}
\DeclareSIUnit\um{\micro\metre}
\DeclareSIUnit\psh{$\si{\hubble}^3 \si{\Mpc}^{-3} \si{\mK}^2$}

\title[Blending impact on EoR detection]{An evaluation of source-blending impact on the calibration of SKA EoR experiments}

\author[C. Shan et al.]{%
Chenxi Shan,$^{1}$\thanks{E-mail: 
 \href{mailto:cxshan@hey.com}{cxshan@hey.com} (CS);
 \href{mailto:hgxu@sjtu.edu.cn}{hgxu@sjtu.edu.cn} (HX)}
Haiguang Xu,$^{1}$\footnotemark[1]
Yongkai Zhu,$^{1}$
Yuanyuan Zhao,$^{1}$
Sarah~V.~White,$^{2}$
Jack~L.~B.~Line,$^{3,4}$
\newauthor
Dongchao Zheng,$^{1}$
Zhenghao Zhu,$^{5}$
Dan Hu,$^{6}$
Zhongli Zhang,$^{5,7}$
and Xiangping Wu$^{8}$
\\
$^{1}$School of Physics \& Astronomy, Shanghai Jiao Tong University, Shanghai, China\\
$^{2}$Department of Physics and Electronics, Rhodes University, PO Box 94, Makhanda, 6140, South Africa\\
$^{3}$International Centre for Radio Astronomy Research, Curtin University, Perth, WA 6102, Australia\\
$^{4}$ARC Centre of Excellence for All Sky Astrophysics in 3 Dimensions (ASTRO-3D)\\
$^{5}$Shanghai Astronomical Observatory, CAS, 80 Nandan Road, Shanghai, China\\
$^{6}$Department of Theoretical Physics and Astrophysics, Faculty of Science, Masaryk University, Kotl\'{a}\v{r}sk\'{a} 2, Brno, 611 37, Czech Republic\\
$^{7}$Key Laboratory of Radio Astronomy and Technology, Chinese Academy of Sciences, 20A Datun Road, Beijing 100012, China\\
$^{8}$National Astronomical Observatories, Chinese Academy of Sciences, 20A Datun Road, Beijing 100012, China\\
}

\date{Accepted 2024 September 08. Received 2024 September 05; in original form 2023 December 30}

\pubyear{2024}

\begin{document}
\label{firstpage}
\pagerange{\pageref{firstpage}--\pageref{lastpage}}
\maketitle

\begin{abstract}
Twenty-one-centimetre signals from the Epoch of Reionization (EoR) are expected to be detected in the low-frequency radio window by the next-generation interferometers, particularly the Square Kilometre Array (SKA). However, precision data analysis pipelines are required to minimize the systematics within an infinitesimal error budget. Consequently, there is a growing need to characterize the sources of errors in EoR analysis. In this study, we identify one such error origin, namely source blending, which is introduced by the overlap of objects in the densely populated observing sky under SKA1-Low’s unprecedented sensitivity and resolution, and evaluate its two-fold impact in both the spatial and frequency domains using a novel hybrid evaluation (HEVAL) pipeline combining end-to-end simulation with an analytic method to mimic EoR analysis pipelines. Sky models corrupted by source blending induce small but severe frequency-dependent calibration errors when coupled with astronomical foregrounds, impeding EoR parameter inference with strong additive residuals in the two-dimensional power spectrum space. 
We report that additive residuals from poor calibration against sky models with blending ratios of 5 and 0.5 per cent significantly contaminate the EoR window. In contrast, the sky model with a 0.05 per cent blending ratio leaves little residual imprint within the EoR window, therefore identifying a blending tolerance at approximately 0.05 per cent. Given that the SKA observing sky is estimated to suffer from an extended level of blending, strategies involving de-blending, frequency-dependent error mitigation, or a combination of both, are required to effectively attenuate the calibration impact of source-blending defects.

\end{abstract}




\section{Introduction}
\label{sec:intro}

The low-frequency radio window opens up new opportunities for 
probing the high-redshift Universe through its unique capability to 
receive the redshifted 21-cm hyperfine line emission of neutral 
hydrogen from the early stages of the Universe, which can probe the 
untethered Cosmic Dawn ($z \sim \numrange{15}{30}$) and the Epoch of 
Reionization (EoR; $z \sim \numrange{6}{15}$) with unprecedented 
precision \citep[see][for reviews]{2006PhRFurlanetto,2012RPPPritchard%
,2013ASSLZaroubi,2016ASSLFurlanetto}. 
The faint nature of the 21-cm signal demands instruments with extremely high sensitivity, wide field-of-view (FoV), and wide-band coverage, making radio interferometers favoured for statistical detection of the 21-cm signal \citep{2020PASPLiu}, presumably in the power spectrum (PS) space, given their flexibility to be designed and constructed as an array that meets these technical demands while also offering precise baseline coverage of EoR-specific scales ($0.1 \lesssim k \lesssim 2$ $\si{\Mpc}^{-1}$).
Major interferometers in the low-frequency band include the Giant Metrewave Radio Telescope (GMRT\footnote{%
  \url{http://www.ncra.tifr.res.in/ncra/gmrt}}; \citealt{1991ASPCSwarup}), 
the Murchison Widefield Array (MWA\footnote{%
  \url{https://www.mwatelescope.org}}; \citealt{2013PASABowman,2013PASATingay}), 
the LOw Frequency ARray (LOFAR\footnote{%
  \url{https://www.astron.nl/telescopes/lofar}}; \citealt{2013A&AvanHaarlem}), 
the MIT Epoch of Reionization (MITEoR\footnote{%
  \url{https://space.mit.edu/home/tegmark/main\_omniscope.html}}; \citealt{2014MNRASZheng}),
the Precision Array for Probing the Epoch of Reionization (PAPER\footnote{%
  \url{http://eor.berkeley.edu}}; \citealt{2010AJParsons}), 
and next-generation arrays, the Hydrogen Epoch of Reionization Array (HERA\footnote{%
  \url{http://reionization.org}}; \citealt{2017PASPDeBoer}) and the Square Kilometre Array (SKA\footnote{%
  \url{https://www.skao.int}}; \citealt{2013ExAMellema,2015askaKoopmans}).
However, owing to the complex baseline designs and complicated instrumental effects of these advanced arrays, \textit{precision} data analysis pipelines are required for EoR experiments to overcome calibration, imaging, and analysis challenges and to minimize exposure to systematics, such as calibration biases \citep{2009ApJDatta,2014MNRASGrobler,2016MNRASWijnholds,2016MNRASPatil,2021MNRASGehlot}, polarization leakages \citep{2017ApJMoore,2017AJHales,2018MNRASDillon,2018MNRASGehlot}, wide-FoV and wide-band imaging errors \citep{2013ApJBhatnagar,2015ASPCJagannathan,2016AJRau,2022MNRASYe} and foreground residuals \citep{2012ApJTrott,2016MNRASChapman,2020ApJNasirudin,2021MNRASHothi}. Therefore, it has become clear that EoR experiments must operate under an infinitesimal error budget with rigorous systematic characterizations to detect 21-cm signals successfully \citep[see][for reviews]{2020PASPLiu,2023JApAShaw}. 

Advanced low-frequency interferometers, such as LOFAR and SKA, can reach new 
detection limits that transform the occupation of the observed low-frequency radio 
sky from sparsely distributed, relatively rare sources (e.g. bright radio quasars 
and radio galaxies) to densely distributed populations constituting the majority 
of the extragalactic sky. In particular, the upcoming SKA1-Low is estimated to 
probe extragalactic discrete radio source (EDRS) populations deep into the faint 
sub-\SI{}{\mJy} radio sky ($\sim$ 4400 $\si{\deg}^{-2}$ above \SI{0.1}{\mJy} at 
\SI{150}{\MHz}) and detect the bulk of galaxies with radio power originating from 
star-forming processes, the central active galactic nucleus (AGN), and a composite of 
both \citep{2016A&ARvPadovani}. This increase in sensitivity comes with increasing 
chances of overlap between the surface brightness distributions of sources 
(known as blending), which introduces systematic effects due to contaminated 
measurements \citep[see][for a review]{2021NatRPMelchior}. 
Without proper identification and mitigation strategies, these blended 
measurements will bias the interferometric calibration and, therefore, contribute 
to the overall EoR error budget by introducing imperfections 
during the construction of sky models, which are required to 
overcome line-of-sight (LoS) effects and imperfect instrumental responses 
\citep{2010MNRASLiu,2018ApJLi,2020PASPLiu,2023JApAShaw} by most, if not all, calibration 
strategies of EoR experiments, including sky-based 
\citep{1984ARA&APearson,2009IEEEPRau}, array-based \citep{1982NaturNoordam,1992ExAWieringa}, 
and hybrid calibration methods \citep{2017arXivSievers,2020ApJKern,2020PASAZhang,2021MNRASByrne}. 

Of all the contributions to the tight EoR error budget, calibration errors, 
especially chromatic residual gain errors originating from sky-model defects, 
are of great importance because biases from poor calibration\footnote{%
  We use the term ‘poor calibration’ to describe scenarios where an inaccurate or imperfect calibration is performed, in contrast to a perfect calibration. Poor calibrations result in an ill-calibrated instrument, thereby introducing calibration errors.} couple with 
contamination from strong astronomical foregrounds ($\sim 10^4 - 10^5$ brighter 
than the 21-cm signal), propagate further down the analysis pipeline by leaving 
\textit{additive} residuals in the measurement space, and ultimately impede the 
parameter inference of the EoR signal 
\citep{2010ApJDatta,2016PASATrott,2016MNRASBarry,2017MNRASEwall-Wice,2019ApJByrne,2022MNRASMazumder}. 
Given the propagation nature of calibration errors, quantification of the impact 
of sky-model defects is achieved by mimicking full EoR analysis pipelines 
\citep[see][for an overview on EoR pipelines]{2023PrAHe} and inferring 
the propagated residuals in the measurement space, particularly in the two-dimensional 
(2D) PS space. Although the exact path for establishing the 
calibration propagation effects in the measurement space varies among studies, we 
can primarily sort the existing efforts into two types based on the approach used 
to mimic an EoR analysis pipeline. The first type uses an end-to-end approach 
to simulate propagation effects \citep[e.g.][hereafter B16]{2016MNRASBarry}, 
whereas the second type directly drives the final PS biases via analytic analysis 
\citep[e.g.][hereafter E17]{2017MNRASEwall-Wice}. Despite the different approaches 
adopted, both types of methods have demonstrated that chromatic residual gain errors 
originating from sky-model imperfections severely damage the detectability of the 
21-cm signal by introducing strong contaminations owing to poor calibration (one to two orders 
of magnitude brighter than the EoR signal, E17) and require an extremely small 
tolerance for model defects ($\sim 10^{-5}$ of spectral features, B16). Therefore, 
even though sky models can never be perfect, defects should always be evaluated and 
factored into the overall EoR error budget.

With its unprecedented sensitivity and baseline coverage in the low-frequency window, the SKA1-Low will contribute primary sky models 
for its observing sky \citep{2017PASATrott}. Thus, it is crucial to identify sky-model 
defects for the SKA and properly evaluate the impact of these defects on SKA 
EoR experiments under the telescope's observation specifications and conditions. 
Given the expected observing depth enabled by the SKA1-Low 
\citep[$\sim$ \SI{0.1}{\mJy} at \SI{150}{\MHz},][]{2015askaPrandoni}, source blending 
will undoubtedly introduce one such defect to sky-model constructions of SKA EoR 
experiments and impact the EoR detections via blending-induced calibration errors. 
Since LOFAR, a pathfinder of SKA1-Low with comparable spatial resolution but lower 
sensitivity, has already suffered from source-blending effects to an extended level 
\citep[$\sim$ 3 per cent under a source density $\sim$ 3100 $\si{\deg}^{-2}$,][]{2021A&AKondapally}, 
we expect the fraction of blended sources observed by the SKA1-Low to be at the same 
level or higher than that of the LOFAR observations, but lower than that of the optical 
surveys. 

To date, there is still a lack of investigations that evaluate the impact of source 
blending on interferometric calibration, estimate the blending-induced calibration 
residual error, and identify blending tolerance for SKA sky-model construction.
In this study, we present the first systematic assessment of source blending in the 
low-frequency radio window for the upcoming SKA1-Low, identify a largely two-fold effect of 
blending-induced uncertainties in both the spatial and frequency domains, and 
investigate the impact of frequency-dependent errors originating from calibrations 
against a sky model corrupted by source blending (hereafter blending-corrupted sky model) for SKA EoR experiments. In contrast to the 
simulation and analytic approaches used to infer calibration effects by B16 and E17, 
respectively, we propose a novel hybrid evaluation (HEVAL) approach combining custom 
end-to-end simulations and analytic calibration analysis mimicking SKA EoR experiment 
pipelines: (i) the end-to-end modules of our hybrid evaluation pipeline simulate radio 
maps of the low-frequency radio sky down to the smallest scales resolvable by the 
SKA1-Low ($\sim$ \SI{6}{\arcsec} at 196 MHz for the foregrounds and $\sim$ \SI{0.245}{\cMpc} 
for the EoR signal), introduce sky-model defects into sky-model construction, include 
SKA1-Low instrumental response with visibility synthesis, and reconstruct the radio 
sky with map-making; (ii) the analytic fraction of the HEVAL pipeline adopts the 
widely used logarithmic implementation of the calibration equations 
\citep[][hereafter the PWL method]{1984ARA&APearson,1992ExAWieringa,2010MNRASLiu} 
to mimic the calibration processes of EoR experiments and estimates blending-induced 
calibration errors and propagation biases. By utilizing a pair of sky models, consisting 
of a perfect sky model and a corrupted one, HEVAL decouples the instrumental noise bias 
from the calibration solution and derives the \textit{relative} errors from poor calibration due 
to sky-model defects. The combination of end-to-end simulation and analytic analysis 
enables HEVAL to isolate the impact originating solely from source-blending defects and 
to quantify the blending-induced error budget under realistic SKA observations.

This paper is organized as follows. Section \ref{sec:blending} introduces the 
definition of source blending adopted for this study, identifies its underlying impact 
on sky-model construction, and evaluates the blending ratio expected for the SKA1-Low. 
Section \ref{sec:meth} presents our HEVAL pipeline and the 2D PS measurement space for 
evaluating the impact of source-blending defects on the calibration of SKA EoR 
experiments. We introduce both the key components of our custom end-to-end 
simulation suite and the analytic formalism of blending-induced residual gain error 
and its propagation under a sky-based calibration scheme. In Section \ref{sec:res}, 
we present the propagated residual power in the 2D PS space and identify the blending 
tolerance of the SKA sky-model construction based on the residual power contamination. 
We then discuss the implications of these results for EoR experiments and provide 
potential strategies for mitigating blending-induced bias in Section \ref{sec:dis}. 
Throughout this work, we adopt a flat $\Lambda$CDM cosmology with 
$H_0 = \SI{100}{\hubble} = \SI{67.66}{\km\per\second\per\Mpc}$,
$\Omega_m = 0.3096$, $\Omega_{\Lambda} = 1 - \Omega_m = 0.6904$, $\Omega_b = 0.0489$,
$n_s = 0.9665$, and $\sigma_8 = 0.8102$.

\section{Source blending}
\label{sec:blending}
\begin{figure*}
  \centering
  \includegraphics[width=0.5\textwidth]{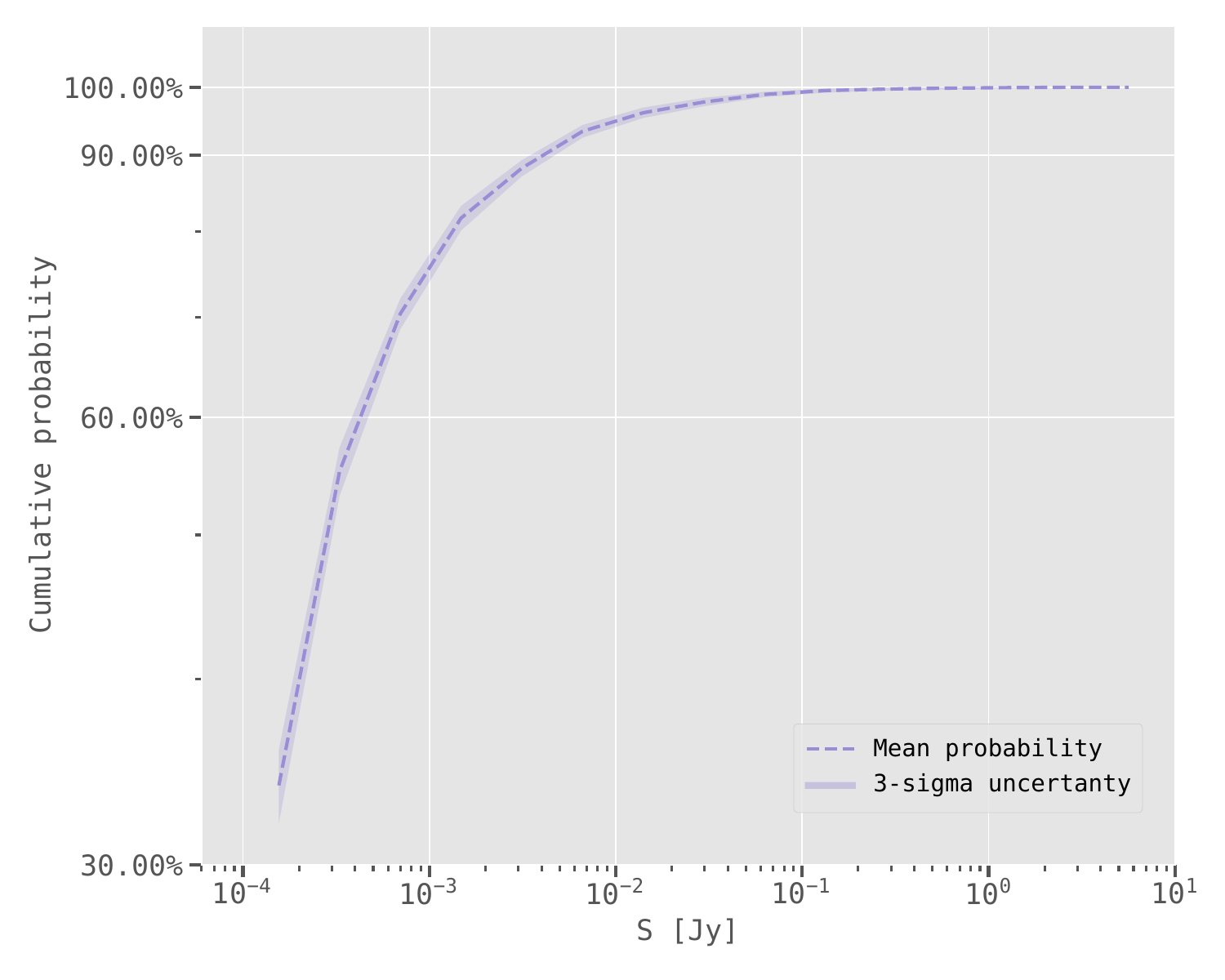}
  \caption{\label{fig:ska_blending_fraction}%
    The cumulative distribution function for blending ratio of extragalactic discrete objects under the SKA1-Low observation specifications and conditions. The x-axis represents the 150 MHz flux density of the discrete radio sources of the SKA1-Low observing sky, plotted on a logarithmic scale in the unit of Jy, whereas the y-axis represents the cumulative probability. The distribution is calculated by dividing the cumulative number of blended sources for each flux-density bin by the total number of blended sources. The shadow marks the 3-sigma uncertainty estimated using a Jackknife resampling technique with 60 subsamples. The lower and upper flux density are limited by the SKA1-Low's detection limit and the W08 simulation, respectively. For this estimation, the SKA observation condition is set using the synthesized beam size (8 arcsec) of the SKA1-Low at 150 MHz; the threshold of blending is set as 0.68; the total blending ratio of all the sources is estimated to be 7.25 per cent. 
  }
\end{figure*}

Blending of astronomical sources often refers to the overlapping of structural 
components in the projected sky. To properly evaluate and simulate the 
source-blending effect, a clear definition of blending is required. In this 
study, \textit{source blending} is defined as independent, unrelated objects (including 
compact sources, elements of multi-component sources, and side lobes) 
distributed close enough in the projected sky plane so that they cannot be 
accurately measured independently under the telescope's resolving condition. 
Based on the spatial relationship between the two objects in the image space 
and the existing blending definitions 
\citep{arcelinDeblendingGalaxiesVariational2020,sanchezEffectsOverlappingSources2021}, 
we identify three different blending cases:
\begin{itemize}
  \item \textbf{Stacked}: The projected radio emission contours of two 
  objects coincide entirely, such that 100 per cent of the flux density 
  distribution overlays each other.
  \item \textbf{Overlapping}: The projected radio emission contours of 
  two objects partially intersect, with greater than 0 per cent but less 
  than 100 per cent of the flux density distribution of one object overlapping 
  the other.
  \item \textbf{Proximity}: The projected radio emission contours of 
  two objects are close to each other but have a 0 per cent 
  overlap in their flux density distribution. The angular separation of 
  the peak flux density centroids can be up to several times the telescope's 
  resolving beam.
\end{itemize}
In the most common form, source blending in the radio domain occurs when two 
individual objects blend into an apparently elongated or reshaped source. 
For simplicity, we only consider the blending of two objects in our study. 
Although we note the existence of multi-source blends, multi-source blending 
cases can be considered as the superposition of two-object blends. Using 
the observation sensitivity estimated by \cite{2015askaPrandoni} and the 
simulated source model from the SKA Designed Study Simulated Skies 
($S^3$\footnote{%
  \url{http://s-cubed.physics.ox.ac.uk}}; \citealt{2008MNRASWilman}, hereafter W08), 
we can roughly estimate the fraction of sources that are blended with other 
sources for upcoming SKA1-Low observations. By calculating the per unit-area 
density of all the sources ($D$, $\si{\arcsec}^{-2}$ ) above the SKA1-Low's 
detection limit and projected extension of each source ($a$, $\si{\arcsec}^2$), 
the likelihood of blending with another source can be inferred as 
$L = min( a / D / T, 1)$, where $T$ ($T \in$ (0, 1]) is the dimensionless 
threshold of blending. Subsequently, a quick and dirty estimation of the 
SKA blending ratio can be achieved by calculating the blending likelihood for 
each source above the SKA1-Low's detection limit. The total blending ratio 
of all the sources for the SKA1-Low is estimated to be approximately $5 - 28$ 
per cent considering different thresholds $T$ and 
projected extensions $a$. As expected, this level of blending falls shy of 
the blending ratio of those optical surveys \citep[see][and references therein]{2021NatRPMelchior}, 
but still poses a significant challenge for SKA sky-model construction. 
In Fig. \ref{fig:ska_blending_fraction}, we present the cumulative distribution 
function of the blending ratio under SKA1-Low's observation specifications and conditions 
at 150 \si{\MHz} with a $T = 0.68$ threshold. Given that the underlying W08 source-count 
models are characterized by Schechter parametrization, blending sources are unsurprisingly 
dominated by sources at the faint end. 

For the sky-model construction, defects due to source blending are introduced 
during the detection and measurement phases. In a typical catalogue-building 
pipeline, source-detection methods decompose radio emission 
into Gaussian components and group the associated components as detected 
sources \citep{2015asclsoftMohan,2018PASAHancock}. The presence of source 
blending often leads to biased detection because the source finder may struggle 
to decompose the correct Gaussian components owing to blended structures. 
In addition, component associations can also lead to source-blending defects 
due to the complex morphology of EDRS populations 
\citep[e.g. see the G4Jy Sample,][]{2020PASAWhiteA,2020PASAWhiteB}. Without 
physical constraints from the host galaxy \citep[links to optical counterparts 
through cross-match techniques, such as][]{2017PASALine}, distinctively separated 
components, such as isolated lobes and nearby faint compact sources, are easily 
associated as one blended source. Therefore, source blending significantly 
affects the accuracy of sky-model construction and can cause uncertainties 
with a largely two-fold impact: (i) \textit{the spatial domain flux density and position error}\footnote{%
  Although there are other errors in the measurement or parameter space (such as angular 
  size and position angle errors) contributed by the mixture of flux density 
  distributions, we consider them alternative descriptions of the same problem.}, 
which originate from a mixture of originally independent flux density distributions, 
and (ii) \textit{the frequency-domain spectral index deviation}, which is caused by 
the blending of intrinsically different spectral components. As most limiting factors of 
sky models (e.g. instrumental noise and defects from sky-model incompleteness) are 
often coupled with one another, methods with the ability to isolate the error contribution 
of one error of origin at a time are required to evaluate the impact of each sky-model 
defect properly. \Cref{sec:meth} introduces one such method to evaluate the 
sole impact of calibration against sky models with source-blending defects for 
SKA EoR experiments.

\section{Methodology}
\label{sec:meth}
\begin{figure*}
  \centering
  \includegraphics[width=1.0\textwidth]{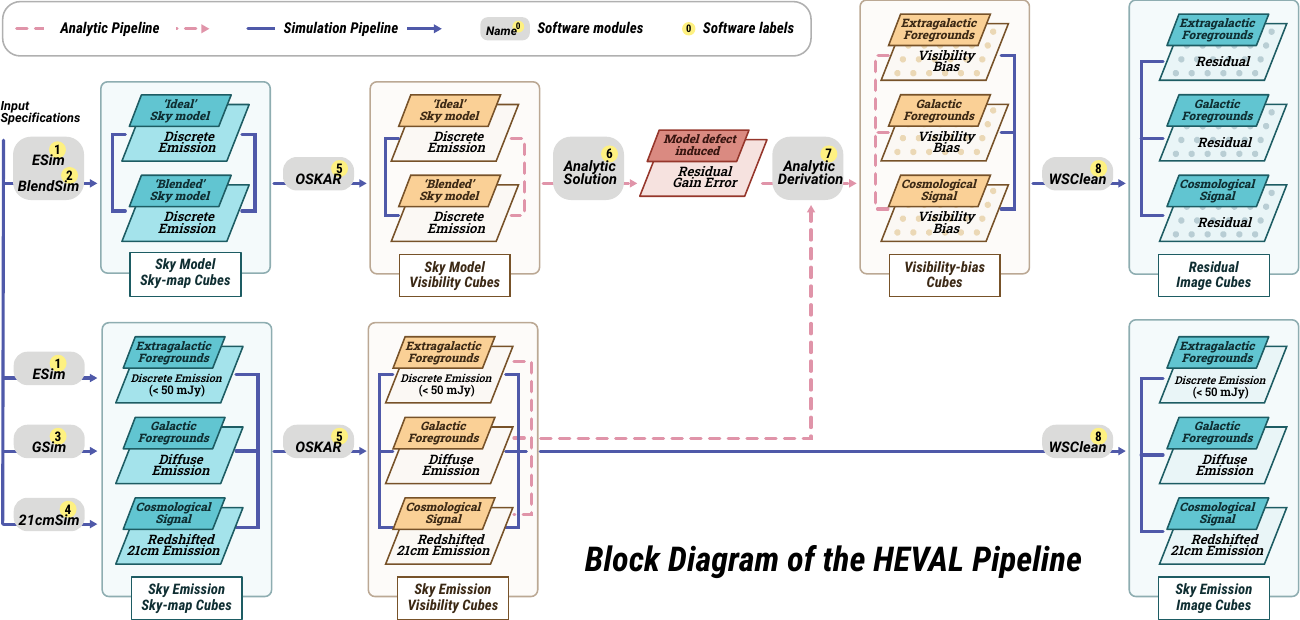}
  \caption{\label{fig:pipeline}%
  	The HEVAL pipeline for sky-model defect evaluation. The solid-blue lines 
  	mark the end-to-end simulation faction for sky-map generation, SKA1-Low visibility 
  	realization, and map making. The dashed pink lines mark the analytic analysis 
  	approach for the gain error estimation and visibility bias calculation.
  	The solid blue lines mark the end-to-end simulation faction, including   	
  	sky-map generation, SKA1-Low visibility realization, and map making, whereas the 
  	dashed pink lines mark the analytic analysis approach for the gain error estimation 
  	and visibility bias calculation. Although typical foreground removal is not explicitly 
  	included in the pipeline, this process is implicitly handled through our analytical 
  	fraction of the hybrid HEVAL pipeline directly in the visibility space, and the 
  	analytic derivation of visibility biases is analogous to the visibility residuals 
  	produced by pipelines subtracting the foregrounds.
 	Coloured double diamonds and rounded grey squares represent data blocks 
    and software blocks, respectively. 
    From the data flow perspective, the upper and lower fractions denote the ill-calibrated 
    residual and sky emission data flow, respectively.
    The simulation of sky model pairs, using 
    \textsc{ESim} (labelled 1) and \textsc{BlendSim} (labelled 2), is introduced 
    in \Cref{subsec:simskymodelandBlending}; the generation of evaluation data, 
    using \textsc{ESim} (labelled 1), \textsc{GSim} (labelled 3), and 
    \textsc{21cmSim} (labelled 4), are detailed in \Cref{subsec:simevaluation}; 
    the visibility realization, using \textsc{OSKAR} (labelled 5), is described 
    in \Cref{subsec:SKAsimulation}; the analytic analysis of the source-blending 
    impact within the visibility space is introduced in \Cref{subsec:cali}; the 
    step-by-step analytic derivation of gain error (labelled 6) and the analytic 
    calculation of the propagation visibility bias (labelled 7) are  
    detailed in \Cref{app:gain-error} and \Cref{app:prop-error}, 
    respectively; and the convolution and deconvolution of the radio maps, using 
    \textsc{WSClean} (labelled 8), are described in \Cref{subsec:CLEAN}; 
    assumptions adopted by the HEVAL pipeline are organized in \Cref{tab:assumptions2}; 
    data-related mathematical notions are noted in \Cref{tab:datasymbols};
    the array configuration and the simulation specifications used for this study 
    are listed in \Cref{tab:arrayspec,tab:simspec}, respectively.
  }
\end{figure*}

This section presents a hybrid approach for evaluating the calibration impact of 
sky-model defects originating from source blending for future SKA EoR experiments. 
Fig. \ref{fig:pipeline} outlines the data and software blocks of our HEVAL pipeline, 
in which the solid blue and dashed pink lines mark the end-to-end and analytic data 
flows, respectively. We introduce the complete set of HEVAL modules\footnote{%
  Careful readers may notice the absence of foreground removal from the HEVAL 
  pipeline. This is due to the fact that no explicit foreground removal is performed by HEVAL. 
  Instead, the ‘foreground-removed’ visibility residuals are derived within the analytic portion 
  of the HEVAL pipeline via analytic estimation.}, including 
sky-model building, visibility synthesis, calibration error estimation, and map-making, 
and eventually arrive at the measurement space to quantify the blending-induced impact 
through 2D PS analysis. 
To better put our method and results into context, 
we attempt to summarize the assumptions adopted by the HEVAL pipeline in 
\Cref{tab:assumptions2}. We list the data-related mathematical notions in 
\Cref{tab:datasymbols} and present the space transformation of the data cubes in 
Fig. \ref{fig:SpaceTransform}.

Firstly, the HEVAL pipeline starts with the simulation in the image space by 
presenting the construction of sky models along with the introduction of 
sky-model defects, resulting in a pair of sky models, one with source-blending 
defects and one without, and the simulation of the low-frequency radio sky, 
which includes extragalactic discrete foregrounds, Galactic diffuse 
foregrounds, and 21-cm signals. All maps are simulated across 
frequency channels forming sky-map cubes [$S_{\rm ori}(l,m,\Delta f)$] along 
the frequency axis. 
The modules used are from our custom \textsc{Fg21Sim+}\footnote{%
  \textsc{Fg21Sim+}: \url{https://github.com/Fg21Sim/Fg21SimPlus}} 
simulation suite, 
which is developed based on our previous simulation 
efforts \citep{2010ApJWang,2013ApJWang,2019ApJLi}, notably the 
\textsc{fg21sim}\footnote{%
  \textsc{fg21sim}: \url{https://github.com/Fg21Sim/fg21sim}} 
simulator. As our next-generation simulation suite, \textsc{Fg21Sim+} is 
designed to simulate the low-frequency radio sky in the image space down to 
the arcsec level to cover the smallest angular scales resolved by high spatial 
resolution instruments, such as LOFAR and SKA1-Low. In addition, \textsc{Fg21Sim+} 
can introduce imperfections into the simulation, making it particularly useful 
for evaluating sky-model defects within data analysis pipelines. 

Secondly, we detail image-to-visibility space transitions within the HEVAL 
pipeline. Cross-correlation synthesis with the SKA1-Low array configuration is 
achieved by feeding the \textsc{Fg21Sim+}-generated sky-map cubes 
of the sky-model pairs, foreground emissions, and the EoR signals
to the \textsc{OSKAR} array simulator, where the output visibility cubes mimic realistic 
instrumental responses under the SKA1-Low observation conditions. The array 
configuration and corresponding SKA1-Low specifications are summarized in 
\Cref{tab:arrayspec,tab:simspec}, respectively. 
Within the visibility space, we estimate the propagation effects during 
interferometric calibration under the sky-based calibration scheme by presenting 
an analytic analysis to infer the \textit{relative} residual gain errors from poor calibration 
owing to sky-model defects and their subsequent propagation visibility biases. 
We use visibilities of the simulated sky model pairs to evaluate the 
blending-induced per-frequency per-antenna relative residual gain error 
and, subsequently, apply the relative error solutions to the visibilities of each 
sky component to infer the per-frequency per-baseline propagation visibility biases. 
Although no explicit foreground-removal module is implemented by HEVAL, the derived 
visibility biases are similar to the residual visibilities generated by pipelines 
explicitly performing foreground removal, owing to the nature of the relative calibration. 
Detailed analytic derivations are introduced in \Cref{subsec:cali}.
In the visibility space, HEVAL produces two sets of visibility cubes: 
(i) the sky emission visibility cubes [$V_{\rm ori}(u,v,\Delta f)$] 
representing the true response of the extragalactic discrete foregrounds, 
Galactic diffuse foregrounds, and EoR signals and (ii) the blending-induced 
visibility-bias cubes [$V_{\rm res}(u,v,\Delta f)$] for each sky component.

Thirdly, the HEVAL pipeline converts the visibilities back to the image space to 
reconstruct the radio sky through the map-making processes presented in 
\Cref{subsec:CLEAN}, given our aim of inferring the residual PS impact under the 
‘reconstructed’-sky approach \citep{2019MNRASMorales}. The \textsc{WSClean} imager is used for the Fourier 
transform, convolution, and de-convolution of the simulated visibilities. Both sets 
of visibility cubes [$V_{\rm ori}(u,v,\Delta f)$ and $V_{\rm res}(u,v,\Delta f)$] are 
converted into two sets of image cubes [$I_{\rm ori}(l,m,\Delta f)$ and $I_{\rm res}(l,m,\Delta f)$] 
for each sky component, respectively.

Finally, we establish the PS measurement space for statistical EoR detection 
and transform the two sets of image cubes into two sets of three-dimensional (3D) PS 
[$P_{\rm ori}(k_x, k_y, k_z)$ and $P_{\rm res}(k_x, k_y, k_z)$] and calculate 
the cylindrical-averaged 2D PS [$P_{\rm ori}(\kperp, \klos)$ and $P_{\rm res}(\kperp, \klos)$] 
for each sky component, respectively. Being a key EoR figure of merit, 2D PS, 
along with EoR windows, is used to evaluate the eventual impact of blending-induced 
poor calibration. 

Throughout this paper, we denote the \textsc{Fg21Sim+}-generated 
sky emission, \textsc{OSKAR}-simulated cross-correlations, \textsc{WSClean}-convoluted, 
and -deconvoluted images as sky maps, ‘observed’ visibilities, dirty maps, and clean 
maps, respectively. 
To clearly distinguish between the terms, we label the terms of sky emission with 
‘original’ (e.g. ‘original’ image cubes and ‘original’ visibility cubes). In contrast, 
we label all the poor-calibration related terms with ‘residual’ (e.g. ‘residual’ visibility-bias 
cubes and ‘residual’ powers). 

\begin{table*}
\begin{threeparttable}
 \caption{Summary of assumptions adopted by the HEVAL pipeline.}
 \label{tab:assumptions2}
 \begin{tabular}{llll}
  \hline
  Assumptions & Pipeline module & Label\tnote{a} & Section\tnote{b}\\[2pt]
  \hline
  Sky models only include EDRS populations & sky-model construction & 1, 2 & \Cref{subsec:simskymodelandBlending} \\[2pt]
  Sky-model defects originated only from blending & sky-model construction & 1, 2 & \Cref{subsec:simskymodelandBlending} \\[2pt]
  Restriction to two-object blending & sky-model construction & 2 & \Cref{subsec:simskymodelandBlending} \\[2pt]
  No noise impact introduced by blending & sky-model construction & 2 & \Cref{subsec:simskymodelandBlending} \\[2pt]
  Physical scales from degree to arcsec level & sky-map generation & 1, 3, 4 & \Cref{subsec:simevaluation} \\[2pt]
  Sky-based per-frequency per-antenna calibration scheme & calibration & 6, 7 & \Cref{subsec:cali} \\[2pt]
  Unpolarized logarithmic approximation & calibration & 6, 7 & \Cref{subsec:cali} \\[2pt]
  Calibrate the amplitude and phase separately & calibration & 6, 7 & \Cref{subsec:cali} \\[2pt]
  Wideband deconvolution with multi-frequency weighting to mitigate spectral artefacts & map-making & 8 & \Cref{subsec:CLEAN} \\[2pt]
  \hline
 \end{tabular}
 \begin{tablenotes}
 \small
 \item[a] The ‘label’ column marks the labelled modules in Fig. \ref{fig:pipeline}.
 \item[b] The ‘section’ column links to the section where the related modules are introduced in the text.
 \end{tablenotes}
\end{threeparttable}
\end{table*}

\begin{table*}
\begin{threeparttable}
 \caption{Data-related mathematical notions used in the paper.}
 \label{tab:datasymbols}
 \begin{tabular}{ll}
  \hline
  Notion & Meaning\\
  \hline
  $\Delta f$ & bandwidth of a frequency band\\[2pt] 
  $l, m$ & direction cosines define the sky coordinates\\[2pt] 
  $u, v$ & spatial frequencies define the antenna spacing coordinates\\[2pt] 
  $k_x, k_y, k_z$ & the 3D wavenumbers\\[2pt] 
  $\kperp, \klos$ & angular and line-of-sight wavenumbers\\[2pt] 
  $V_{\rm ori}(u,v,\Delta f)$, $V_{\rm ori,\,\mathit{A}}(u,v,\Delta f)$\tnote{a} & ‘original’ visibility cube transformed from its corresponding ‘original’ sky-map cube\\[2pt]
  $V_{\rm res}(u,v,\Delta f)$, $V_{\rm res,\,\mathit{A}}(u,v,\Delta f)$\tnote{a} & ‘residual’ visibility-bias cube derived using its corresponding ‘original’ sky-map cube\\[2pt]
  $I_{\rm ori}(l,m,\Delta f)$, $I_{\rm ori,\,\mathit{A}}(l,m,\Delta f)$\tnote{a} &  ‘original’ image cube transformed from its corresponding ‘original’ visibility cube\\[2pt]
  $I_{\rm res}(l,m,\Delta f)$, $I_{\rm res,\,\mathit{A}}(l,m,\Delta f)$\tnote{a} &  ‘residual’ image cube transformed from its corresponding ‘residual’ visibility-bias cube\\[2pt]
  $P_{\rm ori}(k_x, k_y, k_z)$, $P_{\rm ori,\,\mathit{A}}(k_x, k_y, k_z)$\tnote{a} &  ‘original’ PS cube calculated from ‘original’ image cube\\[2pt]
  $P_{\rm res}(k_x, k_y, k_z)$, $P_{\rm res,\,\mathit{A}}(k_x, k_y, k_z)$\tnote{a} &  ‘residual’ PS cube calculated from ‘residual’ image cube\\[2pt]
  $P_{\rm ori}(\kperp, \klos)$, $P_{\rm ori,\,\mathit{A}}(\kperp, \klos)$\tnote{a} &  ‘original’ 2D PS cylindrically averaged from ‘original’ PS cube\\[2pt]
  $P_{\rm res}(\kperp, \klos)$, $P_{\rm res,\,\mathit{A}}(\kperp, \klos)$\tnote{a} & ‘residual’ 2D PS cylindrically averaged from ‘residual’ PS cube\\[2pt]
  $R_{\rm res,\,\mathit{A}/ori,\,\mathit{B}}(\kperp, \klos)$ &  2D PS ratio between the ‘residual’ power of component A and the ‘original’ power of component B\\[2pt]
  $R_{\rm res,\,\mathit{A}/res,\,\mathit{B}}(\kperp, \klos)$ &  2D PS ratio between the ‘residual’ power of component A and the ‘residual’ power of component B\\[2pt]
  \hline
 \end{tabular}
 \begin{tablenotes}
 \small
 \item[a] The two terms mark the general notion and the specific notion of sky component A, respectively.
 \end{tablenotes}
\end{threeparttable}
\end{table*}

\subsection{Simulation of sky models and source-blending effect}
\label{subsec:simskymodelandBlending}

We detail the realization of sky-model construction in the image space. This study follows 
the convention and includes only EDRS populations in the sky model using the \textsc{ESim} module, 
because most commonly used sky models for interferometric calibration are constructed using 
EDRS populations, often from high-quality EDRS catalogues 
\citep{2016MNRASCarroll,2017PASAProcopio,2020PASAWhiteA,2020PASAWhiteB,2021A&AKondapally} and dedicated surveys 
\citep{2011PASANorris,2014MNRASLane,2015PASAWayth,2015A&AHeald,2016A&AJackson,2016ApJZheng,2016AASLacy,2017A&AShimwell,2017A&AIntema,2022PASAHurley-Walker}. 
While diffuse emission contributes power to the radio sky at larger angular scales, 
the limited existing data and algorithmic challenges leave discrete sources as the 
primary ingredients for modelling the sky \citep{2020PASPLiu}.

Our next-generation \textsc{Fg21Sim+} suite offers the flexibility to add defects to 
sky models in the image space through defect-simulation modules, such as 
\textsc{BlendSim}, which introduces additional source-blending effects to the 
\textsc{ESim}-simulated sky models. By combining \textsc{ESim} and \textsc{BlendSim}, 
our simulation of sky-model construction and blending-induced defects arrives 
at a pair of sky models: (i) an ‘\textit{ideal}’ sky model representing the true 
flux density distribution of discrete sources without any blending of sources and 
(ii) a ‘\textit{blended}’ counterpart with some of the sources blended together 
resulting in flux density distribution deviations with respect to (w.r.t) their 
‘\textit{ideal}’ counterparts. The data flow is reflected in the upper-left part 
of Fig. \ref{fig:pipeline}. With the ‘\textit{ideal}’ and ‘\textit{blended}’ 
sky models, our HEVAL pipeline can utilize a paired sky model estimation approach 
to derive the sole contribution of source blending during poor calibration, which is 
achieved by isolating the impact of blending defects from any other limiting 
factors of sky-model construction (e.g. baseline noises and other sky-model defects) 
and inferring the blending-induced relative residual gain errors.
Given that the relative residual gain error and propagation bias are subject to 
the sky model differences in the visibility space, the simulation of sky models 
with end-to-end control and clearly defined procedures is crucial for 
estimating the source-blending effect properly. 

\subsubsection{Simulation of EDRS populations}
\label{subsubsec:ESim}

EDRS populations \citep[see][for reviews]{2016A&ARvPadovani,2019NatAsPanessa,%
2019ARA&ABlandford,2021A&ARvODea} are key objects of interest in the 
radio window, not only due to their diversity in terms of morphological 
types but also due to the underlying physical processes, black-hole accretion 
and star formation, which govern how galaxies evolve. As the bulk of the 
extragalactic sky, EDRSs dominate the radio sky in terms of both number and 
brightness, making them a key foreground component of EoR 
experiments. Consequently, EDRS populations are critical for calibration, imaging, 
and subtraction in the field of 21-cm science. 
In light of this, we compose our sky model simulation by considering only extragalactic radio 
sources with radio emission originating from supermassive black hole (SMBH) 
activity, which comprises compact radio core emission and structural 
features (such as relativistic jets, plumes, lobes, and hotspots), and star 
formation processes, which produce synchrotron radiation through 
relativistic plasma within the supernova remnants, associated with galaxies. 

\begin{figure*}
  \centering
  \includegraphics[width=0.85\textwidth]{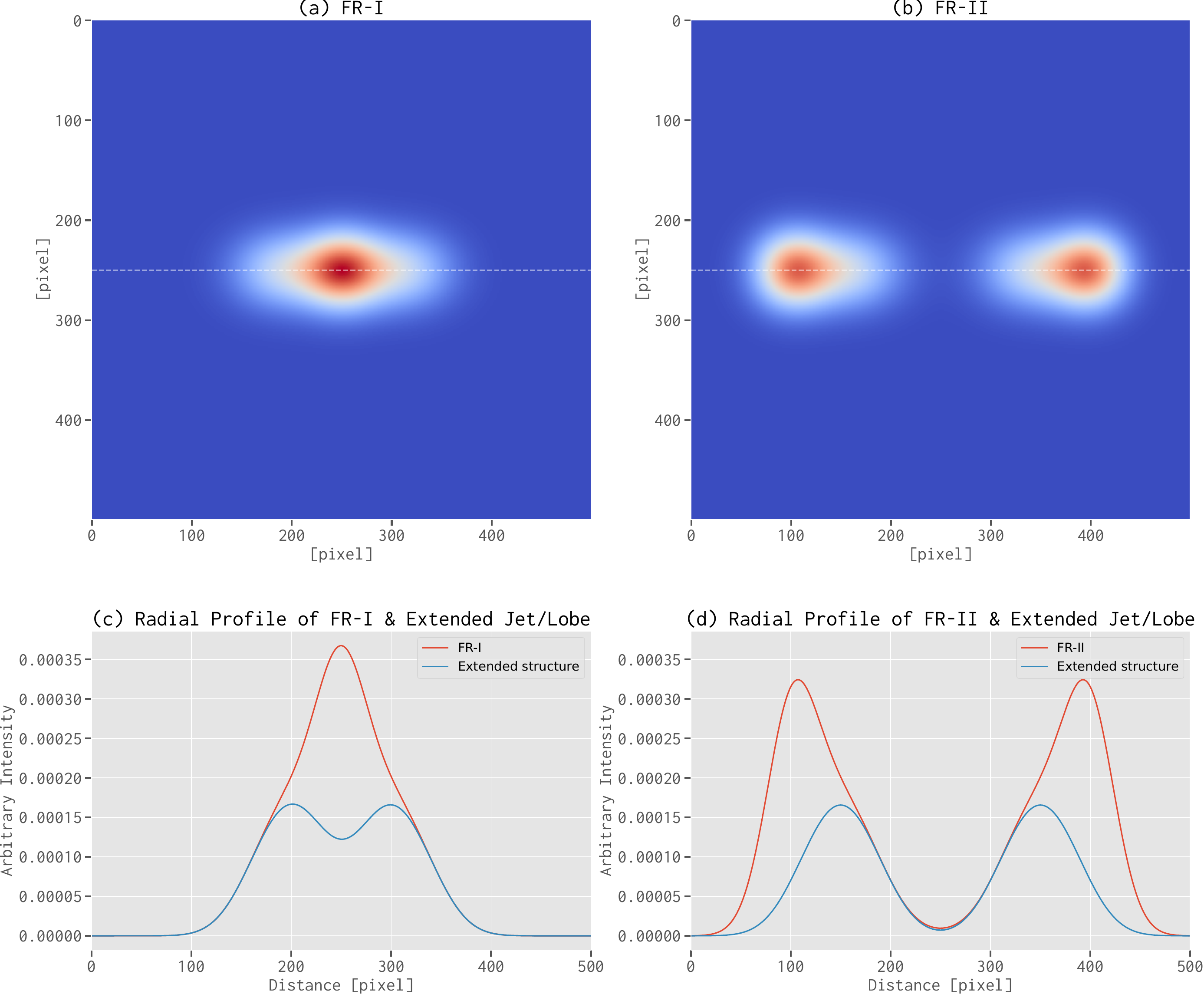}
  \caption{\label{fig:ExtendedSources}%
    Illustration of \textsc{ESim}-generated radio galaxies using a multiple-Gaussian model. (Upper) \textsc{ESim}-simulated flux-density distributions of typical (a) FR-I and (b) FR-II sources. The two flux density distributions are presented using the same colour bar. The dashed white line marks the jet axis. (Lower) The corresponding radial profiles along the jet axis of the two sources above. The blue line marks the radial profile of the extended structures, whereas the red line marks the overall radial profile of the whole source. 
    The two figures from the same column are plotted with the matching pixel coordinates.
  }
\end{figure*}

The EDRS populations included in the \textsc{ESim} module consist of (i) 
bright extended radio galaxies (RGs), which are Fanaroff--Riley (FR) type I 
and type II galaxies \citep{1974MNRASFanaroff}, (ii) compact radio-loud (RL) 
AGNs, which consist of both compact steep-spectrum (CSS) and peaked-spectrum 
(PS) AGNs, (iii) radio-quiet (RQ) AGNs, and (iv) star-forming (SF) and starburst (SB) galaxies. The 
underlying source count distribution follows the commonly used W08 source 
catalogue simulation. To generate sky maps of individual sources, we 
simulated the brightness distribution of compact sources (i.e. compact AGNs, 
SF, and SB galaxies) with a single 2D Gaussian model and extended sources (i.e. FR-I and FR-II RGs) with 
multiple-Gaussian models by adopting morphological properties from 
the W08 catalogue (e.g. angular size and position angle). In particular, the 
extended sources are created from the convolution of a 2D Gaussian kernel with 
a radial profile along the jet axis, which results in a superposition of multiple Gaussian 
distributions, representing the extended structures, such as the jet and lobes of 
radio galaxies (Fig. \ref{fig:ExtendedSources}). For the spectral models, 
\textsc{ESim} utilizes a power-law spectrum for most radio spectra, 
whereas a curvature model from \cite{2000MNRASJarvis} and spectral turnover 
discovered by \cite{1998PASPODea} are implemented for compact core 
emission and PS sources, respectively.

\subsubsection{Simulation of the blending effect}
\label{subsubsec:blending}

\begin{figure}
  \includegraphics[width=0.48\textwidth]{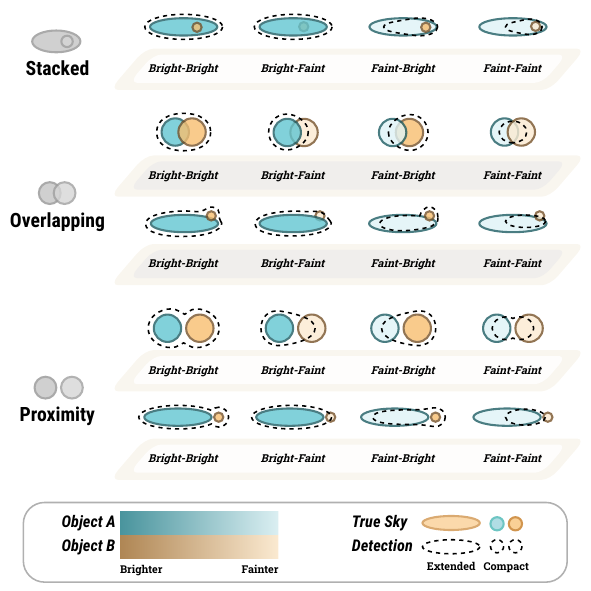} 
  \caption{\label{fig:blendingtypes}
	Most commonly occurred blending types between two components in the image space (the cases of two extended structures fully stacked are omitted due to the rare occurrence). The rule-of-thumb illustrations demonstrate the possible blending-induced error in the image space during source detection with the presence of noises. The elongated and round shapes indicate extended and compact features, respectively. The colour-filled shapes represent the ground truth of the components, whereas the dashed black lines mark the detected components by source finders. For bright sources, the detection errors are dominated by blending. For faint sources, image noises contribute to flux density distribution bias in addition to blending-induced uncertainties. 
  }
\end{figure}

\begin{figure}
  \includegraphics[width=0.45\textwidth]{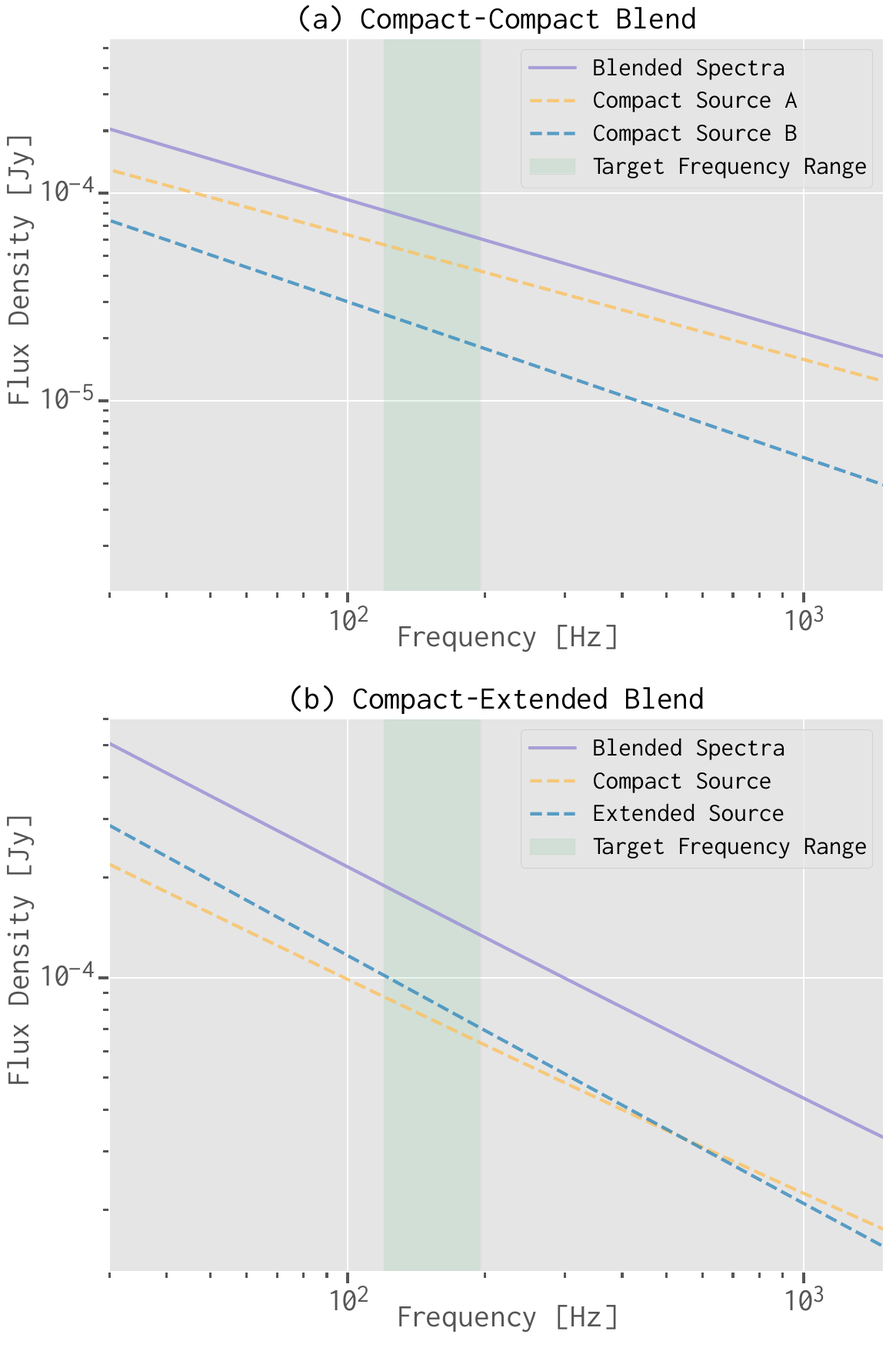}
  \caption{\label{fig:BlendingSpectral}
	Demonstration of blending defects in the spectral domain. (Top) Blended spectra of the two compact radio sources. (Bottom) Blended spectra of compact and extended sources. The two dashed lines mark the original spectra of the individual sources and the purple line marks the spectra of the blended pair. The green shaded area indicates the frequency range used in this study. In both cases, an apparent spectral index deviation exists after the two individual sources are blended.
  }
\end{figure}

The \textsc{BlendSim} module is used to introduce source-blending effects in both the 
spatial (Fig. \ref{fig:blendingtypes}) and spectral domains (Fig. \ref{fig:BlendingSpectral}). 
The implementation of \textsc{BlendSim} contains all the EDRS populations 
in \Cref{subsubsec:ESim}, resulting in 15 different blending types, such 
as RG-RQ AGN, AGN-SF, and RG-SB. To mimic the spatial displacement of brightness 
distribution, a beam with a limited resolution is used to convolute the ideal 
brightness distribution of the selected source pairs in the ‘stacked’, 
‘overlapping’, or ‘proximity’ case by adjusting the distance between 
the centres of two objects. The width of the convolution beam should not be 
larger than that of the SKA synthesized beam, and we set the beam size 
to 6 arcsec for this study. To consider only spatial distribution deviations 
rather than flux density errors, \textsc{BlendSim} employs a blending convolution 
procedure that utilizes a normalized kernel to retain the same total flux 
density of the source pair before and after blending. 

After the source pair is blended in the spatial domain, \textsc{BlendSim} 
adds the blending-induced error in the spectral domain. Because our sky-model building 
does not include spectral-index spatial variations for individual sources, the 
spectral domain error is introduced via the spectral index deviations of the integrated 
spectral indices of the blended pair. Given that most of the blended sources present 
either power-law or power-law-like spectra within our simulated frequency band, 
a power-law\footnote{%
  Unlike spectral fitting with real data, our modelling processes are ideal cases with 
  no noise or errors that may cause spectral fluctuations or distortions. A power-law 
  model approach to mimic spectral index deviation is sufficient for this study.} 
is used to model the integrated spectral index of the blended spectra 
(e.g. the blended spectra presented in Fig. \ref{fig:BlendingSpectral}). 
After the spectral model is assigned, \textsc{BlendSim} generates multi-frequency sky 
models using the ‘\textit{blended}’ monochromic sky model with the estimated blended 
spectral indices.

For the simulation of blending defects, it is important to note that \textsc{BlendSim} 
does not introduce flux density errors, which are most likely unavoidable for blending 
sources in real data considering image noise and artefacts. Flux density errors will not 
only introduce their own impact on calibrations but also bias spectral modelling, leading to 
additional spectral index errors. The coupling of different error sources hinders 
the final interpretation of the impact in the measurement space. Given that our aim is to decouple 
the sole impact of blending, we explicitly choose not to simulate any flux density errors to 
ensure that the evaluation and interpretation target only the source-blending impact. 
While this omission of flux density errors in the simulation does mean that the errors of 
blending are most likely underestimated compared to real-world scenarios, it allows us to 
directly attribute the propagated biases to blending defects alone. This approach provides 
a clean and controlled evaluation, isolating the effects of a single source of origin, without 
the confounding influence of other error sources.

\begin{figure*}
  \centering
  \includegraphics[width=0.85\textwidth]{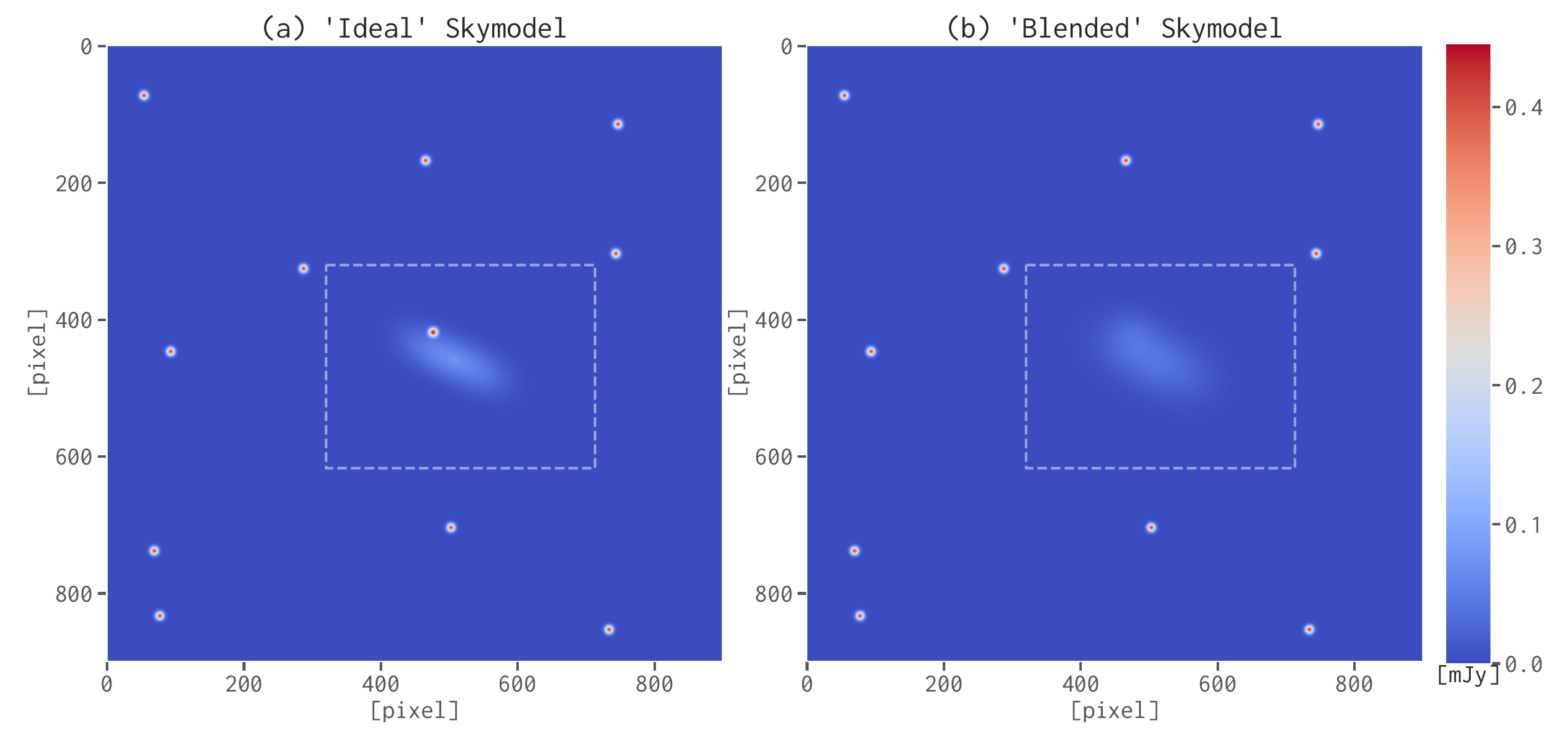}
  \caption{\label{fig:SkyModelPair}%
     Demonstration of a monochromic sky model pair simulated using our \textsc{ESim} and \textsc{BlendSim} models. (Left) A small patch (\smallpatch{1.5}) of an ‘\textit{ideal}’ sky model with the ideal source model pair. (Right) The same sky patch of the corresponding ‘\textit{blended}’ sky model with the blended source model pair. For a sky model pair, the differences between the ‘\textit{blended}’ and ‘\textit{ideal}’ sky models are the flux density distributions of these source model pairs marked by the dashed white lines, whereas the rest of the sources have exactly the same flux density distributions. Since \textsc{BlendSim} normalizes the blending convolution procedure, no flux density uncertainties are introduced for source model pairs. The two figures are plotted using the same colour bar. Multi-frequency sky model pairs are simulated by supplying the monochromatic version with spectral models. For this particular pair, the spectral models were similar to those presented in Fig. \labelcref{fig:BlendingSpectral} (b).
  }
\end{figure*}

\subsubsection{The ‘\textit{ideal}’ and ‘\textit{blended}’ sky model}
\label{subsubsec:skymodelpair}

To generate the sky map of the ‘\textit{ideal}’ sky model and its ‘\textit{blended}’ counterpart, 
a \textit{basis} sample of 4,000 sources, ranging from 0.1 to 1000 mJy, are sampled 
from the W08 catalogue satisfying its underlying source count model with a minimum 
separation requirement of 100 \si{\arcsec} within the estimated SKA1-Low FoV of \SI{5}{\degree}. 
Therefore, the sparsely distributed sources are not biased by blending effects. 
With the sky model \textit{basis} at hand, 
a sub-sample of the \textit{basis} is selected as the host of the blended source pair 
and assigned to another source with a maximum separation of 6 arcsec, 
corresponding to the synthesized beamsize of the SKA1-Low. Using the ideal 
flux density distribution and the true spectral index of both the source 
pair and the \textit{basis} sample, sky-map cubes of the ‘\textit{ideal}’ multi-frequency sky models 
are generated. As for the counterpart, the blended flux density distribution and 
deviated spectral index are simulated using \textsc{BlendSim}. By replacing 
the ideal model of the source pair with the blended model, we obtain sky-map cubes 
of the ‘\textit{blended}’ multi-frequency sky models. In Fig. \ref{fig:SkyModelPair}, we 
present a small patch of a pair of ‘\textit{ideal}’ and ‘\textit{blended}’ sky models. 
Adopting the 15 types of 
blending pairs described in \Cref{subsubsec:blending}, \textsc{BlendSim} assigns the 
types based on the likelihood using the W08 number count model. 

For a systematic estimation of the source-blending impact on the SKA EoR experiments, 
a set of 2, 20, and 200 blended source pairs are generated for the \textit{basis} 
sample of 4,000 sources, representing source-blending levels of 0.05 per cent, 0.5 
per cent, and 5 per cent, respectively. Because we aim to find an expectedly modest 
blending tolerance of sky-model construction for SKA EoR experiments, the 5 per cent 
minimum of the estimated SKA blending ratio is used as the maximum of the three levels.
In the latter part of the manuscript, we refer to the three blending levels as ‘mild’, 
‘moderate’, and ‘high’ blending ratios, respectively. For simplicity, sky models 
affected by the three blending ratio levels are marked as ‘mildly’-, ‘moderately’-, 
and ‘highly’-corrupted sky models, respectively.

\subsection{Simulation of the evaluation data}
\label{subsec:simevaluation}

\begin{figure*}
  \centering
  \includegraphics[width=0.8\textwidth]{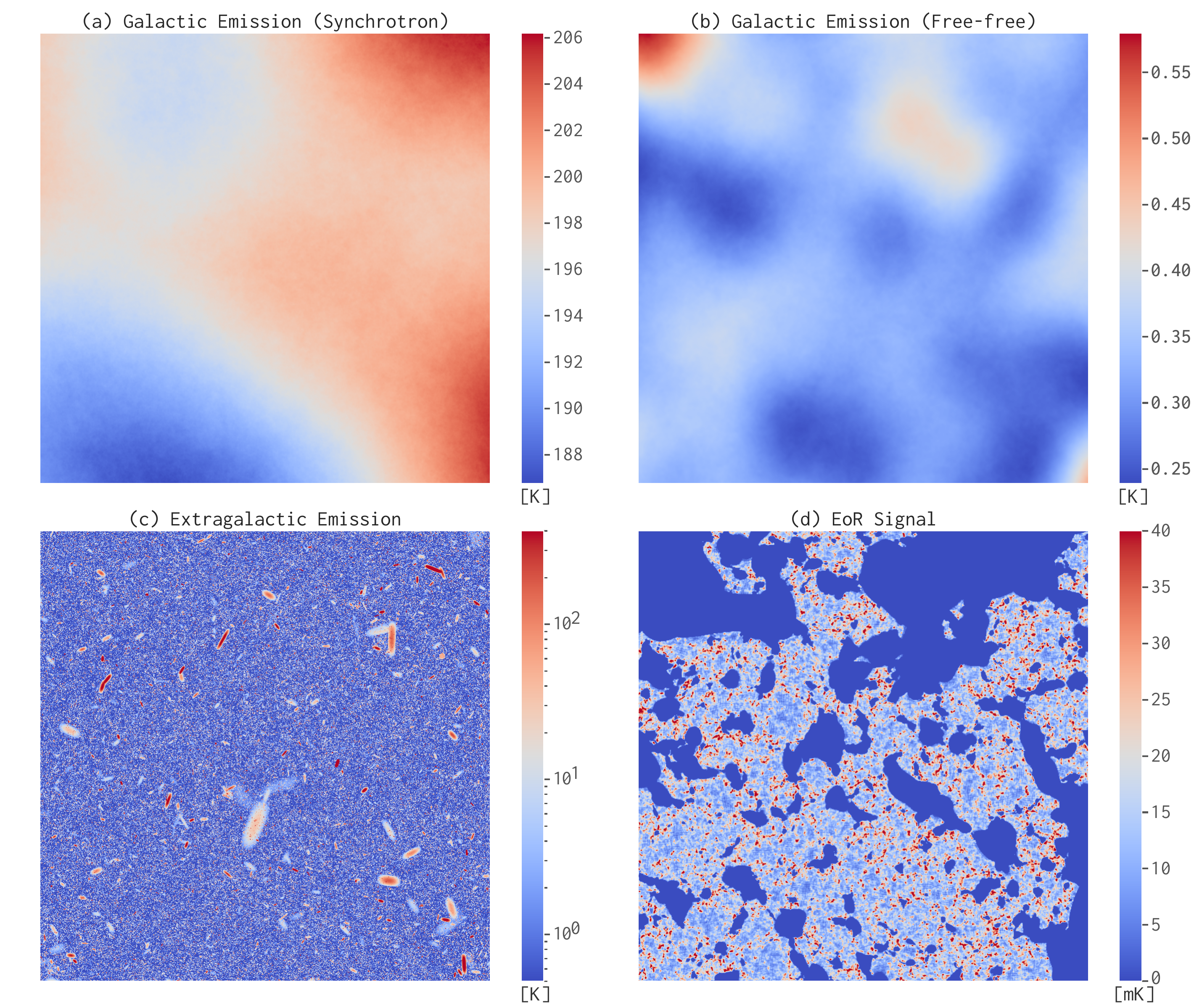}
  \caption{\label{fig:EvaluationData}%
     Simulated sky maps of evaluation data representing the low-frequency radio sky: the Galactic diffuse (a) synchrotron and (b) free-free emission, the (c) extragalactic discrete emission, and the (d) EoR signal at \SI{158}{\MHz}. All the images are zoomed in to the central \patch{1} to show the small-scale structures. Each figure uses its own colour bar for the display. In particular, the EoR signal is plotted using unit [mK] different from the rest.
  }
\end{figure*}

Evaluation data cubes are required to represent the low-frequency radio sky, including 
the EoR signal and its foregrounds, to evaluate the propagation effect of residual gain errors. 
The SKA1-Low has baselines across an extensive range of lengths that can probe the
 EoR signal across approximately three decades in scale. The ability to detect the 
 EoR signal at a substantial number of scales, both large and small, gives the edge 
 of the SKA1-Low over other EoR-aimed instruments. 
Even though the total power of the 21-cm signal is expected to be constituted mainly 
from large-scale contributions, 
the EoR signal at the medium and small scales pokes unique aspects of the Universe, 
particularly during the later stages of the EoR. 
For one thing, the 21-cm signal can trace the small-scale structure and its coupling 
to the large-scale structure \citep{2007ApJLidz}. For another, the EoR signal is 
expected to be highly non-Gaussian during the later stages of reionization 
\citep{2016MNRASMondal,2018MNRASMajumdar}, and its non-Gaussianity is probed at the 
smaller spatial scales. 
Therefore, to thoroughly investigate the foreground coupling effect of the blending-induced 
calibration error for SKA EoR experiments, it is crucial to simulate the radio sky containing 
physical scales that are spatially resolvable by the SKA1-Low baselines. 
In the remainder of this section, we introduce the \textsc{Fg21Sim+} modules used to generate 
evaluation data containing simulated scales from the degree level down to the smallest 
scales resolvable by the SKA1-Low. The data flow of the low-frequency radio sky simulation 
is illustrated in the lower left part of Fig. \ref{fig:pipeline}. The corresponding  
\textsc{Fg21Sim+}-generated sky maps are presented in Fig. \ref{fig:EvaluationData}.

\subsubsection{Extragalactic foregrounds}
\label{subsubsec:EDRS}

The extragalactic foregrounds, dominated by EDRS populations (see 
\Cref{subsubsec:ESim}), predominate the contamination of the EoR signal 
mainly on smaller scales because they are discrete sources with flux density 
fluctuations at the arcsec to arcmin scales. Although other extragalactic sources, 
such as diffuse radio emission associated with galaxy clusters, present 
unique challenges to the EoR experiment and impose contamination at the arcmin
scale \citep{2019ApJLi,2022MNRASZhou}, they are ultimately excluded from our 
model because of both the relatively faint flux, which is three to four orders 
of magnitude lower, and a significantly lower number density compared to the EDRS populations. 
We use the \textsc{ESim} module to generate sky maps of the extragalactic foregrounds, 
adopting the simulation specifications from \Cref{subsubsec:spec}. 
Here, the full EDRS source-count model of the W08 catalogue is implemented, 
in contrast to the sampling procedure when simulating sky models detailed in \Cref{subsec:simskymodelandBlending}. 

\subsubsection{Galactic foregrounds}
\label{subsubsec:Galactic}

The Galactic foregrounds mainly contain the diffuse Galactic synchrotron 
and free-free emission, with the synchrotron component being the dominant 
among the two. 
Our \textsc{GSim} treats both the synchrotron and free-free components 
separately and simulates a patch of the sky using basis template maps: 
(i) For Galactic synchrotron emission, the basis synchrotron template map 
deployed by \textsc{GSim} 
is the reprocessed Haslam \SI{408}{\MHz} all-sky map\footnote{%
  The reprocessed Haslam \SI{408}{\MHz} map:
  \url{http://www.jb.man.ac.uk/research/cosmos/haslam_map/}}, 
which is referred to as the HAS14 map, by \cite{2015MNRASRemazeilles} with 
significantly improved removal of both extragalactic sources and instrument 
artefacts. To calculate the sky maps in the SKA1-Low frequency 
via extrapolation with a power-law model from the \SI{408}{\MHz}, the 
all-sky spectral index map ($\alpha \in$ [2.50, 3.20], \citealt{2002A&AGiardino})
of the Galactic synchrotron emission is used. (ii) As for the Galactic free-free 
emission, the basis template map is simulated through 
indirect estimations, as direct observations of the free-free component, 
which is overwhelmed by the strong synchrotron counterpart, is not 
possible. Owing to the common origin of free-free and H$\alpha$ emission, 
a tight relation between the two has been demonstrated. It can be utilized 
to calculate the free-free emission \citep[see][and references therein]{2003MNRASDickinson}. 
The \textsc{GSim} module largely follows the prescription of \cite{2003MNRASDickinson}, 
employs the H$\alpha$ survey from \cite{2003ApJSFinkbeiner}, corrects the 
dust absorption with a full-sky 100-\si{\um} dust map \citep{1998ApJSchlegel}, 
and subsequently, simulates the free-free emission. Because the free-free 
and H$\alpha$ relation is frequency dependent, the conversion from the 
H$\alpha$ map to the free-free map is calculated for the desired SKA1-Low 
frequency window.

Limited by existing observations, Galactic contributions to the radio sky 
are primarily at large scales. Under the scope of this work, flux density 
fluctuations at small scales are also required to properly estimate the 
impact at the longer baselines of the SKA1-Low. Our implementation in 
\textsc{GSim} also includes the addition of Galactic small-scale fluctuations 
using the realization of a Gaussian random field (GRF). Our implementation of adding small-scale fluctuations 
is similar to the existing efforts based on the realization of GRFs 
\citep{2007A&AMiville,2013A&ADelabrouille,2015MNRASRemazeilles,2016MNRASHerv,2017MNRASThorne} 
but with some enhancements. The fundamental idea is to extrapolate the angular PS 
($\mathcal{C}_{\ell}$, parametrized by $\gamma$) to the required small scales ($\ell$) and 
generate the small-scale fluctuations ($G_{\mathrm{ss}}$, parametrized by $\alpha$ and $\beta$) 
using a GRF corresponding to the extrapolated $\mathcal{C}_{\ell}$. We detail the step-by-step 
realization of adding small-scale fluctuations in \Cref{app:smallscales}. 

To generate sky maps of the galactic foregrounds, we set the simulation centre 
of sky maps at a high galactic latitude centre position (R.A., 
Dec.\@ = \SI{0}{\degree}, \SI{-27}{\degree}) since future SKA observations 
are expected to point at high galactic latitudes (e.g. $\abs{b} > \SI{60}{\degree}$) 
to minimize contamination.
As for the small-scale fluctuations, we use the best-fitting of the three required 
parameters: (i) the $\gamma$ index of synchrotron and free-free 
component is fitted as -2.220 and -2.426, respectively, using the full sky template 
maps; (ii) the $\alpha$ and $\beta$ index of the small-scale maps are fitted for the 
synchrotron ($\alpha=0.0342$, $\beta=0.227$) and free-free emission ($\alpha=0.00785$, 
$\beta=0.526$) using a patch of the template map centred at the selected simulation centre. 

\subsubsection{EoR signal}
\label{subsubsec:EoRSignal}

The final component of the radio sky is the 21-cm signal. 
To compensate for the need for high spatial resolution and a relatively large 
FoV, a fast semi-numerical approach is used in the \textsc{21cmSim} module 
by adopting the \textsc{21cmFAST}\footnote{%
  \textsc{21cmFAST}: \url{https://github.com/21cmfast/21cmFAST}}
\citep{2011MNRASMesinger,2020JOSSMurray} package. The cosmic reionization 
process is simulated using the same physical parameters (A. Mesinger, priv. comm.) as the ‘faint galaxies’ 
model of the \emph{Evolution Of 21\,cm Structure} project\footnote{%
  Evolution Of 21\,cm Structure project: \url{http://homepage.sns.it/mesinger/EOS.html}} 
\citep{2016MNRASMesinger} with a box of 294 comoving \si{\Mpc} and 1200 
cells along the side. Our simulation run resolves the smallest physical scale at 0.245 \si{\cMpc}, 
making the finest-resolution 21-cm simulation to date using \textsc{21cmFAST}. From the simulated 
light-cone object, the designated sky maps of each frequency band are sliced w.r.t the corresponding 
redshift and tiled according to the simulation specifications listed in \Cref{tab:simspec} to form sky-map cubes.

\subsection{The realization of simulated SKA observations}
\label{subsec:SKAsimulation}

This section presents the transformation from the image space to the visibility space. 
First, we detail the SKA array configuration used and the corresponding technical 
specifications of the given array. Then, we describe the simulation specifications 
under limited computational powers. Finally, we introduce the realization of visibility 
synthesis using an array simulator which outputs ‘observed’ visibilities with 
instrumental responses of the SKA1-Low. In the end, we transform all the sky-map cubes 
into visibility cubes for both the sky model pairs and the sky components, as 
presented in the middle-left part of Fig. \ref{fig:pipeline}.

\subsubsection{Array configuration}
\label{subsubsec:config}

As the most advanced and transformational project in radio astronomy 
to date, the upcoming SKA will be constructed in two phases, owing to 
challenges and difficulties in designing, constructing, 
and operating extremely large telescopes. Among the two telescopes of 
Phase 1, SKA1-Low in Australia and SKA1-Mid in South Africa, the low-frequency 
aperture array SKA1-Low operating in the \SIrange{50}{350}{\MHz} frequency 
range is the instrument intended for EoR detection and 21-cm cosmology.

Both the SKA1-Low and SKA1-Mid instruments are yet to be fully built, 
although pathfinders, precursors, and verification systems have been built 
and operated as stepping stones for the operation of SKA1. For this study, we use 
the array configuration from the SKA1 System Baseline Design \citep{2013DewdneyBasline,2016DewdneyBasline,2019DewdneyBaseline}. 
These reports detail that SKA1-Low will contain $131072$ log-period dipole 
antennas with band coverage from \SIrange{50}{350}{\MHz}, assembling 512 
stations of diameter $35 - 40$ \si{\m} each consisting of 256 antennas. Among 
all the stations, 224 stations are placed randomly within the ‘core’ 
region of 1000 \si{\m} in diameter, and 288 stations are distributed into 
‘clusters’ along the three array spiral arms, which extend the baselines up to 
65 \si{\km}. Following the SKA1-Low Design Baseline, the estimated SKA1-Low 
specifications are summarized in \Cref{tab:arrayspec}.

\begin{table}
\begin{threeparttable}
\caption{Array specifications adopted for the SKA1-Low.}
\label{tab:arrayspec}
\begin{tabular}{lc}
\hline
Parameters                         & Values \\
\hline
No. of stations                    & 512                        \\
No. of antennas per station        & 256                        \\
Station size                       & 35 m                       \\
Max. baseline                      & 65 km                      \\
Max. spatial resolution\tnote{a}   & $\sim \SI{6}{\arcsec}$ at 196 \si{\MHz} ($\propto \text{\textlambda}$)   \\
Primary beam\tnote{a}              & $\sim \SI{3}{\degree}$ at 196 \si{\MHz} ($\propto \text{\textlambda}$)   \\
Frequency range                    & 50-350 MHz                 \\
Max. bandwidth                     & 300 MHz                    \\
Max. no. of channels               & 55,296                     \\
Max. frequency resolution          & 226 Hz                     \\
\hline                    
\end{tabular}
\begin{tablenotes}
\small
\item[a] The spatial resolution and primary beam are estimated at 196 \si{\MHz}; the rest of the specifications are taken from the Design Baseline reports \citep{2013DewdneyBasline,2016DewdneyBasline,2019DewdneyBaseline}.
\end{tablenotes}
\end{threeparttable}
\end{table}

\subsubsection{Simulation specification estimation}
\label{subsubsec:spec}

The high spatial resolving power of the SKA1-Low advocates the capability 
of SKA EoR experiments to probe the structural scale ranging from several arcsecs 
to degrees via 21-cm signals. 
In accordance with the estimated spatial resolution of the SKA1-Low 
($\sim \SI{6}{\arcsec}$ at 196 \si{\MHz}), the simulated sky maps are 
required to contain the smallest scale at the 6 arcsec level ($\propto \text{\textlambda}$) 
at the corresponding frequency, which 
expects the pixel size of the sky map to be approximately 2 arcsec. Within the 
designed SKA1-Low frequency window, three sparsely selected frequency bands, 
with a bandwidth of 8 MHz are chosen to limit the possible cosmological evolution 
of the EoR signal \citep[e.g.][]{2013ApJThyagarajan,2019ApJLi}, 
covering 120 -- 128 \si{\MHz}, 154 -- 162 \si{\MHz}, and 188 -- 196 \si{\MHz}. 

Simulating the radio sky with both high spatial and frequency resolution 
across extensive sky coverage presents significant computational challenges. 
This is because each stage of the process requires substantial resources: 
(i) the cosmological simulation of the 21-cm signal is memory-intensive, 
(ii) the generation of multi-frequency sky maps demands a high-performance GPU, 
and (iii) the visibility synthesis and subsequent map-making processes require 
substantial CPU power. Consequently, the combined high demands on CPU, GPU, and 
memory inherently restrict the practical specifications of our simulations. 
Given these limitations, it is impractical to simulate the SKA observing sky 
in a way that achieves a large field of view (FoV) and high spatial resolution 
simultaneously. Since spatial resolution plays a more critical role under the 
scope of this study (e.g. EDRS-only sky models and determination of blending of 
sources), we prioritize spatial resolution over areal coverage. However, to still 
adequately cover the key EoR scales ($0.1 \lesssim k \lesssim 2$ $\si{\Mpc}^{-1}$), 
we chose a reasonable balance to simulate sky patches of up to \patch{2}. 
The simulation specifications for both sky models and evaluation data are presented in \Cref{tab:simspec}.

\begin{table*}
\begin{threeparttable}
  \caption{%
    Simulation specifications of the sky models and evaluation datasets. 
  }
\label{tab:simspec}
\begin{tabular}{lrrrrrr}
\hline
Component                 & Usage       & Sky Coverage\tnote{a} & Pixel Size\tnote{a}      & Exposure\tnote{a}      & Pre-processing\tnote{b}                 & Post-processing\tnote{c} \\
\hline
‘\textit{Ideal}’ sky model           & Calibration & \patch{5}    & \SI{1}{\arcsec} & \SI{2}{\min}  & None                          & None            \\
‘\textit{Blended}’ sky model         & Calibration & \patch{5}    & \SI{1}{\arcsec} & \SI{2}{\min}  & None                          & None            \\
Extragalactic foregrounds & Evaluation  & \patch{2}    & \SI{2}{\arcsec} & \SI{6}{\hour} & Bright sources are masked     & Clean images are cropped as \patch{1} \\
Galactic foregrounds      & Evaluation  & \patch{2}    & \SI{2}{\arcsec} & \SI{6}{\hour} & Tapered at \SI{1.5}{\degree}  & Clean images are cropped as \patch{1} \\
EoR signal                & Evaluation  & \patch{2}    & \SI{2}{\arcsec} & \SI{6}{\hour} & None                          & Dirty images are cropped as \patch{1} \\
\hline                    
\end{tabular}
\begin{tablenotes}
\small
\item[a] Due to the short exposure times and visibility-only need, the simulation of sky models is less computationally expensive than the evaluation datasets. Therefore, sky models are simulated using a larger sky coverage with a smaller pixel size.
\item[b] Galactic synchrotron and free-free emission are combined as one component. Prior to the generation of ‘observed’ visibilities, the combined Galactic sky maps are tapered to reduce the window effect. 
\item[c] All final images of the evaluation datasets are cropped in order to mitigate the imaging error of the marginal regions due to insufficient CLEANing.
\end{tablenotes}
\end{threeparttable}
\end{table*}

\subsubsection{Array simulator}
\label{subsubsec:OSKAR}

The realization of visibilities with practical instrumental effects of the SKA1-Low is 
archived with the \textsc{OSKAR}\footnote{%
  \textsc{OSKAR}: \url{https://github.com/OxfordSKA/OSKAR} (v2.8.3)} 
array simulator \citep{2010Mort} by generating SKA1-Low ‘observed’ visibilities 
with the SKA1-Low specification and configuration using GPU accelerations. 
The specific input telescope model is created using the SKA1-Low 
array layout with antenna coordinates\footnote{%
  \raggedright SKA1-Low Configuration Coordinates:
  \url{https://astronomers.skatelescope.org/wp-content/uploads/2016/09/SKA-TEL-SKO-0000422_02_SKA1_LowConfigurationCoordinates-1.pdf}
  (released on 2016 May 31)} 
from \cite{2016DewdneyCoordinates}. The three frequency bands (\Cref{subsubsec:spec}) 
are divided into 51 channels with a channel width of \SI{160}{\kilo\hertz}. 
To minimize the pointing effect, we choose to centre all the sky maps at the sky position of (R.A., Dec.\@) = (\SI{0}{\degree}, \SI{-27}{\degree}), which is the expected 
SKA1-Low zenith.

Sky-map cubes of both the ‘\textit{ideal}’ and ‘\textit{blended}’ sky models are directly 
passed to the \textsc{OSKAR} simulator without pre-processing processes under a sky coverage 
of \patch{5} and a pixel size of 1 arcsec with a single 2-min snapshot exposure time. 
The sky-map cubes of the evaluation data, extragalactic discrete emission, 
Galactic diffuse emission, and EoR signal, are simulated with a sky coverage of \patch{2} 
and a pixel size of 2 arcsec with deep observations for \SI{6}{\hour}. 
For the EoR foregrounds, pre-processing progresses are made before the visibility synthesis. 
Being extended in nature, the Galactic components, synchrotron 
and free-free, are combined and tapered, which reduces the window effect of 
sky maps containing large-scale structures with a limited size before they 
are ‘observed’ by the simulator. Since the brightest sources of EDRS are already 
known to strongly bias the EoR observations \citep[e.g.][]{2010ApJDatta}, we assume 
these sources are adequately dealt with and removed during data analysis 
pipelines. Hence, our simulations mask the EDRS populations with a \SI{158}{\MHz} 
flux density above \SI{50}{\mJy}, as indicated in Fig. \ref{fig:pipeline}. 
All the specifications and pre- and post-processing processes are listed in \Cref{tab:simspec}.

\subsection{Evaluation of blending-induced calibration errors}
\label{subsec:cali}

This section introduces the analytic fraction of the HEAVL pipeline, which is 
shown in the central part of Fig. \ref{fig:pipeline}. 
A sky-based per-frequency per-antenna calibration scheme is first introduced as 
the basic framework. We then present the derivation of the frequency-dependent 
per-antenna gain error caused by a blending-corrupted sky model w.r.t the 
‘\textit{ideal}’ sky model counterpart through analytic analysis based on the 
PWL logarithmic method. Unlike traditional 
methods aimed at a direct antenna-based calibration solver to infer the 
solution of the complex gain factors \citep{1982RaScThompson}, 
we use an alternative approach to estimate the blending-induced relative gain error 
using a pair of ‘\textit{ideal}’ and ‘\textit{blended}’ sky models. 
By adopting the PWL logarithmic implementation, the analytic expressions arrive at 
two simple sets of linear systems of equations, resulting in an 
overdetermined linear regression problem mathematically. 
To solve for the estimated gain error, we use a singular value decomposition (SVD)-based method. 
With the solved relative residual gain error, we can derive an analytic expression of 
the propagation visibility bias of the sky signal in the visibility space induced by 
an ill-calibrated instrument. 
For conciseness, we state the critical steps in this section and leave 
the step-by-step mathematical derivation in \Cref{app:full-derivation}. A full description 
of the calibration-related mathematical notations is presented in \Cref{tab:allsymbols}.

\subsubsection{Calibration in a sky-based scheme}
\label{subsubsec:skycali}

Considering an array with $N$ antennas, the $i$th antenna measures the 
unpolarized voltage $v_i$ at the antenna end at a single frequency 
channel $f$ within a limited time span $\Delta t$, and can be estimated as
\begin{equation}
  v_i = g_i s_i {+} n_i, \label{eqn:defination}
\end{equation}
where $g_i$ is the antenna complex gain factor, $n_i$ denotes the intrinsic 
antenna noise, and $s_i$ represents the true sky signal. As interferometers 
measure cross-correlations, each baseline correlates measurements from a pair of 
antennas, resulting in a time-averaged visibility of the baseline $\bm{{B}}_{ij}$ 
in the form of
\begin{subequations}\label{eqn:correlation}
\begin{align}
  V_{ij}& \equiv \left\langle v_i^*v_j \right\rangle \label{eqn:correlation-a}\\
  &= g_i^{*}g_j \left\langle s_i^*s_j \right\rangle + g_i^{*} \left\langle s_i^*n_j \right\rangle + g_j \left\langle n_i^*s_j \right\rangle + \left\langle n_i^*n_j \right\rangle  \label{eqn:correlation-b}\\
  &\approxeq g_i^{*}g_j \left\langle s_i^*s_j \right\rangle + n_{ij}. \label{eqn:correlation-c}
\end{align}
\end{subequations}
By assuming that only sky radio signals are correlated, the last three terms 
of equation (\ref{eqn:correlation-b}) can be reduced to noise $n_{ij}$ specific to 
the baseline $\bm{{B}}_{ij}$. Since each antenna gain is a complex factor, 
it can be parametrized by its amplitude $\eta$ and phase $\phi$ as
\begin{equation}
  g_i \equiv e^{\eta_i + \mathrm{i}\phi_i}. \label{eqn:complex-gain}
\end{equation}
If we use the parametric form of each antenna complex gain, the baseline 
measurement question, equation (\ref{eqn:correlation-c}), takes a new form:
\begin{equation}
  V_{ij} = \mathrm{exp} \left[ \left( \eta_i + \eta_j \right) + \mathrm{i} \left( \phi_j -\phi_i \right) \right] \left\langle s_i^*s_j \right\rangle + n_{ij}. \label{eqn:visibility-form2}
\end{equation}

Ideally, the calibration process solves both the true complex gain factor 
$g_i$ of each antenna and the true sky signal cross-correlation 
$S_{ij} \equiv \left\langle s_i^*s_j \right\rangle$ of each baseline
simultaneously. An array of $N$ antennas measures $C^2_N$ visibilities, 
whereas the variables to solve for are $N$ true antenna gain amplitude 
$\eta_i$ items, $N$ true antenna phase $\phi_i$ items, and $C^2_N \equiv N(N-1)/2$ 
true sky cross-correlations $S_{ij}$. Therefore, the measurements are insufficient 
to directly solve for $G_{ij} \equiv g_i^{*}g_j$ and recover $S_{ij}$. However, 
by adopting a sky-based calibration scheme, which supplies a sky model 
to replace the unknown sky cross-correlation, the calibration of $G_{ij}$ becomes 
an overdetermined problem that can be solved using a least-squares solution. 
The sky cross-correlations, $S_{ij}$, can then be recovered by 
applying the solution of the estimated gain values. Calibrating all frequencies 
separately can recover a set of sky cross-correlations within the observed band.

\subsubsection{Analytic analysis of the calibration error}
\label{subsubsec:cali-error}

As stated in \Cref{sec:intro}, sky-based calibration techniques are ultimately 
restricted by the supplied sky model. Consequently, the precision and fidelity 
of these models becomes critical. Source blending, along with various limiting 
factors that couple with one another [such as model completeness 
\citep[e.g.][]{2016MNRASBarry,2021MNRASGehlot}, 
polarized emission \citep[e.g.][]{2017ApJMoore,2018MNRASAsad}, 
and diffuse components \citep[e.g.][]{2022ApJSLanman,2022MNRASSims}], 
collectively impacts the accuracy of calibration processes and subsequently dictates 
the detection ability of the EoR signal. 
To individually evaluate the source-blending impact on the overall 
calibration process, we propose a theoretical estimation that confines blending 
defects as the only sky model limiting factor using a pair of sky models consisting 
of an ideal sky model and its blended counterpart. 
The sky model pairs include only discrete sources and their simulation realization is 
presented in \Cref{subsec:simskymodelandBlending}. Under this paired sky model estimation 
approach, we can use the measured visibility equation (equation \ref{eqn:visibility-form2}) to infer 
the per-frequency per-antenna gain error induced by a ‘\textit{blended}’ sky model w.r.t its 
‘\textit{ideal}’ counterpart within the sky-based calibration scheme. We refer to this type of 
calibration error as the relative residual gain error.

To begin with, by using an ‘\textit{ideal}’ sky model $V^{\mathrm{ideal}}$ to calibrate the $N$ 
antenna composed array, we have the measured visibility and true gain amplitudes 
and phases ($\eta$ and $\phi$, respectively) for the baseline $\bm{{B}}_{ij}$ in 
the form of 
\begin{equation}
V_{ij} = \mathrm{exp} \left[ \left( \eta_i + \eta_j \right) + \mathrm{i} \left( \phi_j -\phi_i \right) \right] V_{ij}^{\mathrm{ideal}} + n_{ij}. \label{eqn:visibility-ideal}
\end{equation}
Then, by supplying a ‘\textit{blended}’ sky model $\acute{V}^{\mathrm{blend}}$, the same 
baseline measurement, albeit ill-calibrated, will take the form:
\begin{equation}
V_{ij} = \mathrm{exp} \left[ \left( \acute{\eta}_i + \acute{\eta}_j \right) + \mathrm{i} \left( \acute{\phi}_j -\acute{\phi}_i \right) \right] \acute{V}_{ij}^{\mathrm{blend}} + n_{ij}, \label{eqn:visibility-blend}
\end{equation}
where $\acute{\eta}$ and $\acute{\phi}$ are the gain factors a ‘\textit{blended}’ 
sky model solves. Subsequently, by combining equations (\ref{eqn:visibility-ideal}) with  (\ref{eqn:visibility-blend}) under the PWL logarithmic method (see the 
step-by-step derivation in \Cref{app:gain-error}), we can obtain two sets of linear 
equations with the form
\begin{subequations}\label{eqn:comb-vis-pair}
\begin{align}
\Delta \eta_i + \Delta \eta_j &= R_{ij}, \label{eqn:comb-vis-real}\\
\Delta \phi_j - \Delta \phi_i &= I_{ij}, \label{eqn:comb-vis-imag}
\end{align}
\end{subequations}
where both the relative gain errors ($\Delta \eta_i \equiv \eta_i - \acute{\eta}_i $ and $\Delta \phi_i \equiv \phi_i - \acute{\phi}_i$, respectively) and the visibility difference in terms of amplitude and phase 
($R_{ij} \equiv \ln \abs{\acute{V}_{ij}^{\mathrm{blend}}} - \ln \abs{V_{ij}^{\mathrm{ideal}}}$ 
and $I_{ij} \equiv \arg \abs{\acute{V}_{ij}^{\mathrm{blend}}} - \arg \abs{V_{ij}^{\mathrm{ideal}}}$, respectively) 
are substituted for simplicity.
For each baseline, both $R_{ij}$ and $I_{ij}$ are only subject to the visibility 
difference between the pair of sky models. Because we supplied simulated sky model 
pairs with source blending as the only limiting factor, the error contribution of 
a ‘\textit{blended}’ sky model w.r.t an ‘\textit{ideal}’ counterpart can be inferred by solving 
the amplitude and phase differences separately with those two simple sets of equations. 
In this way, we decouple the source-blending impact on the calibration of EoR experiments 
from both the baseline noises and other limiting factors of sky-model constructions.

Although recent efforts have been made to demonstrate solving the complex 
gain factor as a complex optimization for polarization and direction-dependent 
calibration \citep[e.g.][]{2015MNRASSmirnov,2018MNRASGrobler}, we adopt the conventional 
approximation and treat the real and imaginary parts separately for our unpolarized case. 
Therefore, we can determine the blending-induced 
calibration error individually by solving the two overdetermined systems of equations and derive 
the per-frequency $\Delta \eta_i$ and $\Delta \phi_i$ for each antenna. A Python-based SVD solver 
is implemented to derive the solution following the exact derivation presented in \Cref{app:error-sol}. 
Because there is no contribution of baseline noise or 
possible biases from specific antenna-based calibration solvers, the solutions 
should be considered as a theoretical upper-limit estimation of the gain errors. 

Finally, by supplying the per-frequency per-antenna relative residual gain errors back 
to equation (\ref{eqn:visibility-form2}) for each sky component (we note true visibility $S^{\mathrm{true}}_{ij}$ 
and ill-calibrated visibility $\acute{S}^{\mathrm{ill}}_{ij}$), we can infer the per-frequency per-baseline 
propagation bias of the sky signal in the visibility space as:
\begin{subequations}\label{eqn:bias}
\begin{align}
\Delta \mathrm{H}_{ij} &= \left[ \mathrm{exp} \left( \Delta \eta_i + \Delta \eta_j \right) - 1 \right] \abs{S_{ij}^{\mathrm{true}}} , \label{eqn:bias-sky-real}\\
\Delta \Phi_{ij} &= \Delta \phi_j - \Delta \phi_i, \label{eqn:bias-sky-imag}
\end{align}
\end{subequations}
where $\Delta \mathrm{H}_{ij} \equiv \abs{{\acute{S}_{ij}^{\mathrm{ill}}}} - \abs{S_{ij}^{\mathrm{true}}}$ 
and $\Delta \Phi_{ij} \equiv \arg \abs{\acute{S}_{ij}^{\mathrm{ill}}} - \arg \abs{S_{ij}^{\mathrm{true}}}$ 
denote the amplitude and phase bias, respectively. 
Based on equation (\ref{eqn:bias}), the derived amplitude and phase biases can be viewed as residuals 
after the removal of the corresponding foreground components without explicitly performing foreground subtraction. 
Because we use the true visibility of the foreground, the visibility bias is \textit{additive} in nature, even after 
a \textit{perfect} foreground removal. The detailed derivation is presented in \Cref{app:prop-error}. 
By utilizing the relative gain errors that are independently estimated for each frequency, the propagation bias 
can be inferred for all channels within each observation band. Thus, with the input visibility cubes 
[$V_{\rm ori,\,gal}(u,v,\Delta f)$, $V_{\rm ori,\,ext}(u,v,\Delta f)$, and $V_{\rm ori,\,eor}(u,v,\Delta f)$], 
we can explicitly derive the visibility-bias cubes for each sky component, namely, 
$V_{\rm res,\,gal}(u,v,\Delta f)$, $V_{\rm res,\,ext}(u,v,\Delta f)$, and $V_{\rm res,\,eor}(u,v,\Delta f)$, respectively.

\subsection{Deconvolution and imaging}
\label{subsec:CLEAN}

As the final fraction of the HEVAL pipeline, this section details map-making, 
which is a crucial final step to enter the measurement space under the 
‘reconstructed’-sky approach. To deal with the challenge of both wideband and 
wide-FOV imaging, the HEVAL pipeline uses the \textsc{WSClean}\footnote{%
  \textsc{WSClean}: \url{https://sourceforge.net/p/wsclean} (v2.6)}
imager \citep{2014MNRASOffringa} tailored for map-making in the low-frequency 
radio sky. 

To reconstruct 3D representations of the radio sky, image cubes along an 
observation frequency band are required. This is achieved 
by converting the 3D visibility cubes back to the image space. Both the 
‘original’ visibility cubes (the lower part of Fig. \ref{fig:pipeline}) 
and the propagated ‘residual’ visibility-bias cubes (the upper part of Fig. \ref{fig:pipeline}) 
of each sky component are converted directly into ‘original’ [$I_{\rm ori}(l,m,\Delta f)$] 
and ‘residual’ image cubes [$I_{\rm res}(l,m,\Delta f)$], respectively. 
In particular, the direct imaging of foreground visibility-bias cubes, instead of 
ill-calibrated visibility cubes, allows an efficient map-making phase, given that 
imaging fidelity in the spatial domain is less a priority than spectral fidelity. 
Because we aim to evaluate the impact of frequency-dependent calibration errors, 
any spectral variations during the deconvolution phase need to be dealt with. 
The wideband deconvolution mode with multi-frequency weighting of the \textsc{WSClean} 
\citep{2017MNRASOffringa} under a Briggs robustness weighting set to $-0.5$ \citep{1995PhDTBriggs} 
is used to account for spectral variations by griding the multi-frequency imaging weights 
together and cleaning all frequency channels jointly. 
In addition, a polynomial is fitted to the full bandwidth, reducing possible imaging noise 
and spectral artefacts that are typically caused by cleaning false peaks impersonated by side 
lobes or noise peaks. These measures allow HEVAL-produced foreground image cubes to mitigate 
possible spectral fluctuations, which might be coupled with the effect of frequency-dependent 
calibration errors in the measurement space. For the 21-cm signal, dirty images are used 
directly for analysis due to the insufficient deconvolution effect of the CLEAN 
algorithm to diffuse faint signals. All the final images are cropped to retain only the 
central \patch{1} regions to mitigate imaging errors due to insufficient CLEANing of the 
marginal regions.

\subsection{The 2D power spectrum \& EoR window}
\label{subsec:eor}

\begin{figure*}
  \centering
  \includegraphics[width=0.79\textwidth]{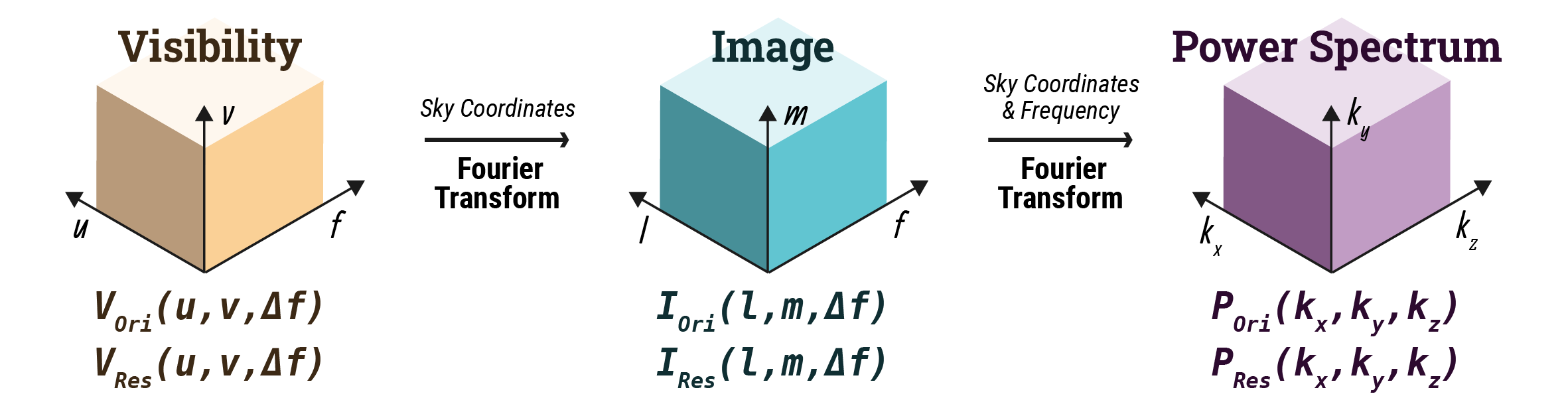}
  \caption{\label{fig:SpaceTransform}%
     The data flow of our ‘reconstructed’-sky PS analysis pipeline. The multi-frequency visibilities of a given band are transformed into image cubes reconstructing the sky emission through the Fourier transform in the sky coordinates. Subsequently, the reconstructed image cubes are transformed into the measurement space through the Fourier transform in both the sky coordinates and the frequency dimension to analyse the spatial features of each sky component. Both ‘original’ visibility and ‘residual’ visibility-bias cubes for each sky component are transformed following the data flow. The $V_{\rm ori}$  ‘original’ visibility cubes are converted to $I_{\rm ori}$ and eventually transform into $P_{\rm ori}$. The $V_{\rm res}$  ‘residual’ visibility-bias cubes follow the same trajectory to the image and PS space as  $I_{\rm res}$ and $P_{\rm res}$, respectively. This plot is inspired by fig. 2 of \protect\cite{2004ApJMorales}.
  }
\end{figure*}

As a redshifted line emission, observations of the EoR 
signal within a frequency span are three-dimensional, including two 
spatial dimensions across the sky and one vertical frequency dimension to the 
LoS. Statistical detection of the EoR signal under the ‘reconstructed’-sky 
approach is expected to be achieved using the 3D nature of the EoR observation 
by converting reconstructed image cubes to 3D Fourier representations [$P(k_x, k_y, k_z)$] in the 
measurement space via Fourier transform. The data flow from interferometric 
observations to the PS measurement space is presented in Fig. \ref{fig:SpaceTransform}.

Within the measurement space, PS analysis can be achieved by averaging $P(k_x, k_y, k_z)$  
either in spherical shells of radii $k$, 
due to its spherical symmetry, resulting in the 1D PS $P(k)$, or over the 
angular annuli of radii $\kperp \equiv \sqrt{k_x^2 + k_y^2}$ for each 
LoS plane $\klos \equiv k_z$, owing to the independence between the spatial 
and frequency dimensions, yielding the 2D PS $P(\kperp, \klos)$. Theoretically, 
the cylindrical-averaged $P(\kperp, \klos)$ confines spectrally-smoothed foregrounds  
to be distributed within the low-\klos{} region of the $(\kperp, \klos)$ plane, 
leaving the rest of the space relatively free of foreground contamination. Thus, the 2D 
PS analysis is widely adopted as the most crucial figure of merit for 
experiments aimed to detect the EoR signal.

However, the reality of complicated instrumental and observational effects, 
such as chromatic primary beams, redistributes the foreground power to higher 
\klos dimensions via mode mixing, resulting in an expanded wedge-shaped 
region known as the foreground wedge \citep{2010ApJDatta,2012ApJMorales,2014ApJPober}. 
Beyond the foreground wedge, a region known as the EoR window \citep[for more details 
on the topic, please refer to][]{2014PhRvDLiu,2014PhRvDDillon}, 
is relatively free of foreground power contamination and can be located at 
\begin{equation}
  \label{eq:eor-window}
  \klos \geq \frac{H(z) D_{\!M}(z)}{(1+z) c} \left[
    \kperp \sin\Theta + \frac{2\pi w f_{\mathrm{21-rf}}}{(1+z) D_{\!M}(z) B} \right]
\end{equation}
\citep{2013ApJThyagarajan,2019ApJLi},
where $H(z)$ denotes the Hubble parameter at redshift $z$, $D_{\!M}(z)$ 
measures the transverse comoving distance, $c$ is the speed of light, $B$ 
is the bandwidth of the frequency window, $f_{\mathrm{21-rf}}$ denotes 
the frequency of the 21-cm line emission in the rest-frame, 
$z = f_{\mathrm{21-rf}}/f_c - 1$ is the signal redshift of the central 
frequency $f_c$ within the measured frequency band, $w$ is the number 
of characteristic convolution widths of the spillover region due to frequency 
response variations, and $\Theta$ is the foreground source angular separation 
from the primary beam centre. Based on the angular separation $\Theta$, commonly 
used EoR window boundaries include ‘FoV’- and ‘horizon’-bound windows, 
which are defined by supplying the FoV and horizon of the instrument. 

Frequency-dependent calibration errors due to source blending are expected to 
aggravate the effect of mode mixing for ‘reconstructed’-sky PS, which will cause further 
foreground contamination by leaving additive ‘residual’ powers within EoR 
windows. Thus, to fully evaluate the impact of source blending on 
the EoR experiments, we discuss the 3D imprint (both the spatial and spectral deviations) 
of the blending-induced visibility bias with the ‘original’ [$P_{\rm ori}(\kperp, \klos)$] 
and ‘residual’ 2D PS [$P_{\rm res}(\kperp, \klos)$] for each sky component, which is calculated 
using the ‘original’ [$I_{\rm ori}(l,m,\Delta f)$] and ‘residual’ image cubes 
[$I_{\rm res}(l,m,\Delta f)$], respectively. 

\section{Results}
\label{sec:res}

\subsection{‘Residual’ powers of EoR foregrounds}
\label{res:ResPowerFG}
\begin{figure*}
  \centering
  \includegraphics[width=1.0\textwidth]{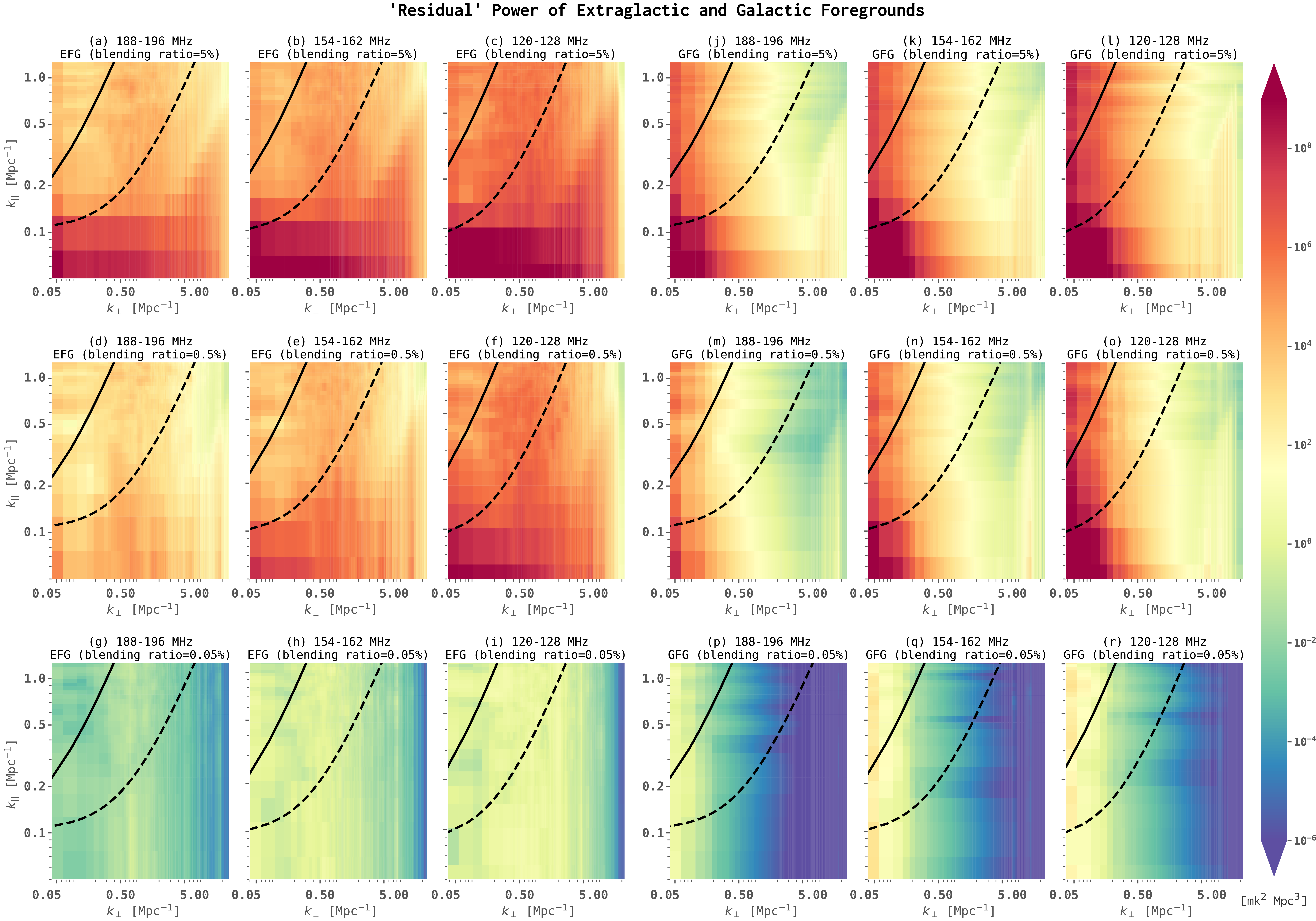}
  \caption{\label{fig:res:Respower}%
    (Left, a-i) Extragalactic ‘residual’ powers $P_{\rm res,ext}(\kperp, \klos)$ and (right, j-i) Galactic ‘residual’ powers $P_{\rm res,gal}(\kperp, \klos)$ originating from blending-induced calibration errors. Within each 3$\times$3 grid, the three rows mark the ‘high’ (5 per cent), ‘moderate’ (0.5 per cent), and ‘mild’ (0.05 per cent) blending scenarios, respectively; the three columns mark the 188 -- 196 MHz, 154 -- 162 MHz, and 120 -- 128 MHz frequency band, respectively. Setting $w$ as 2, the black lines define the EoR windows with the dashed and solid lines marking the ‘FoV’ and the ‘horizon’ boundary, respectively. All the plots share the same logarithmic colour bar using unit [$\si{\mK}^2 \si{\Mpc}^{3}$]. 
  }
\end{figure*}

We calculate the cylindrical-averaged 2D PS using image cubes 
produced by the HEVAL pipeline. In Fig. \ref{fig:res:Respower}, we present the 
‘residual’ 2D PS [$P_{\rm res,ext}(\kperp, \klos)$ and $P_{\rm res,gal}(\kperp, \klos)$] 
transformed from foreground ‘residual’ image cubes [$I_{\rm res,ext}(l,m,\Delta f)$ and 
$I_{\rm res,gal}(l,m,\Delta f)$] for each component owing to the propagation effects 
of blending-induced calibration errors. Given that these powers will still be present 
in the PS space even after the perfect removal of spectrally-smoothed foregrounds, 
we refer to them as ‘residual’ powers. ‘Residual’ powers originating from the ‘highly’-corrupted 
sky model (5 per cent blending ratio) show a substantial occupation of the low $\klos$ modes 
for both the Galactic and extragalactic foregrounds. The strongest Galactic ‘residual’ 
powers reside at larger spatial scales ($\kperp \lesssim 0.2$), while the strong 
extragalactic ‘residual’ powers spread across the large and intermediate spatial scales. For the 
high $\klos$ modes, we can see a clear excess of powers within the EoR window above both 
the dashed black ‘FoV’ line and solid black ‘horizon’ line. These ‘residual’ powers 
contaminate the expected foreground-free regions. Hence, they pose a threat to the parameter 
inference of the EoR signal. For the ‘moderate’ case (0.5 per cent blending ratio), the 
‘residual’ powers, albeit weaker, distribute similarly to that of the ‘high’ blending ratio 
case in the 2D PS space. Unlike the prior two cases, the ‘mildly’-corrupted sky model 
(0.05 per cent blending ratio) introduces 
much weaker ‘residual’ powers with a slightly different vertical distribution, as there is no 
characteristic distribution of ‘residual’ powers occupying low $\klos$ modes for both the Galactic 
and extragalactic foregrounds. However, the contamination of the ‘residual’ power within the EoR 
window remains. 

\begin{figure*}
  \centering
  \includegraphics[width=0.8\textwidth]{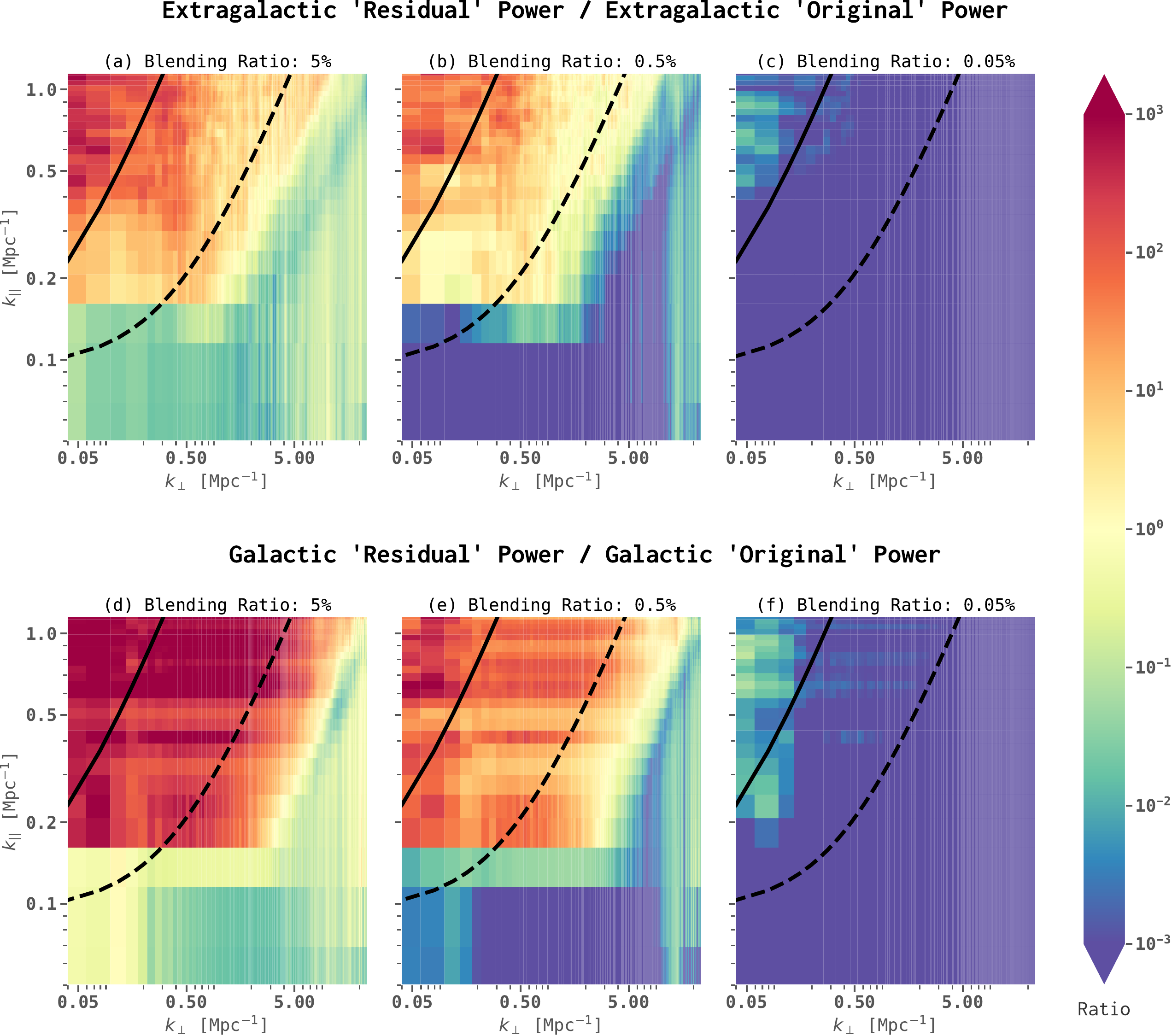}
  \caption{\label{fig:res:RespowerRatio}%
    The 154 -- 162 \si{\MHz} 2D PS ratio between foreground ‘residual’ and ‘original’ powers $R_{\rm res,fg/ori,fg}(\kperp, \klos)$. (Top) $R_{\rm res,ext/ori,ext}(\kperp, \klos)$ ratios under (a) ‘high’ (5 per cent), (b) ‘moderate’ (0.5 per cent), and (c) ‘mild’ (0.05 per cent) blending scenarios. (Bottom) Corresponding $R_{\rm res,gal/ori,gal}(\kperp, \klos)$ ratios under (d) ‘high’ (5 per cent), (e) ‘moderate’ (0.5 per cent), and (f) ‘mild’ (0.05 per cent) blending scenarios. 
    Setting $w$ as 2, the black lines define the EoR windows with the dashed and solid lines marking the ‘FoV’ and the ‘horizon’ boundary, respectively.
    Figs. \labelcref{fig:res:RespowerRatio,fig:res:RespowerEvG,fig:res:ResRatioEoR,%
    fig:res:RespowerEvE,fig:res:weighting1,fig:res:weighting2} share the same logarithmic colour bar, showing only six orders of magnitude for easy viewing.
  }
\end{figure*}

\begin{figure*}
  \centering
  \includegraphics[width=0.8\textwidth]{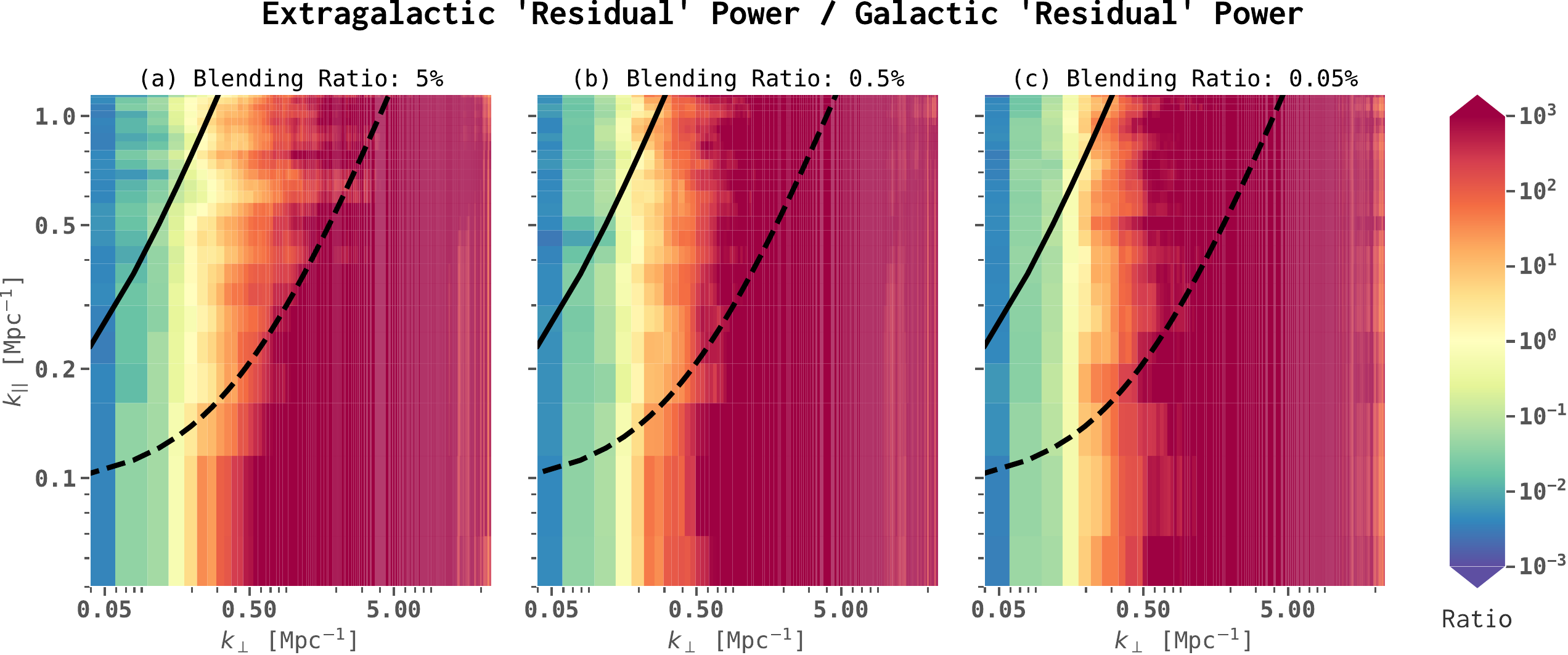}
  \caption{\label{fig:res:RespowerEvG}%
    The 154 -- 162 \si{\MHz} 2D PS ratios between extragalactic and Galactic ‘residual’ powers $R_{\rm res,ext/res,gal}(\kperp, \klos)$ under (a) ‘high’ (5 per cent), (b) ‘moderate’ (0.5 per cent), and (c) ‘mild’ (0.05 per cent) blending scenarios.
    Setting $w$ as 2, the black lines define the EoR windows with the dashed and solid lines marking the ‘FoV’ and the ‘horizon’ boundary, respectively.
    Figs. \labelcref{fig:res:RespowerRatio,fig:res:RespowerEvG,fig:res:ResRatioEoR,%
    fig:res:RespowerEvE,fig:res:weighting1,fig:res:weighting2} share the same logarithmic colour bar, showing only six orders of magnitude for easy viewing.
  }
\end{figure*}

By taking the 2D PS ratio between the ill-calibrated ‘residual’ power 
[$P_{\rm res,ext}(\kperp, \klos)$ and $P_{\rm res,gal}(\kperp, \klos)$] and ‘original’ 
sky emission power [$P_{\rm ori,ext}(\kperp, \klos)$ and $P_{\rm ori,gal}(\kperp, \klos)$] 
for each foreground component, we can identify the region with excess power caused by 
blending-induced calibration errors. Given the nature of frequency-dependent calibration 
errors that cause relatively small fluctuations across the frequency channels, the region 
with excess power should be located at higher $\klos$ modes. In Fig. \ref{fig:res:RespowerRatio}, 
we can see excess powers locate at the region covering important high $\klos$ modes ($\klos \gtrsim 0.2$), 
as expected. Among the three cases, the ‘mild’ blending case has the lowest excess power. 
All the ‘residual’ powers introduced by the ‘mildly’-corrupted sky model are below 10 per cent 
of the ‘original’ sky emission powers. In contrast, the other two cases pose significant 
contamination by introducing excess foreground ‘residual’ powers, which are 10 to over 1000 times 
the ‘original’ sky emission powers, covering most of the EoR window. Outside the 
‘FoV’ window, foreground ‘original’ powers still dominate the wedge region for 
all three cases. Given that we treat the two foreground components separately, we can also 
calculate the 2D PS ratio between the extragalactic and Galactic ‘residual’ 
powers to infer the dominant modes of each foreground component. Fig. \ref{fig:res:RespowerEvG} 
illustrates the spatial scales that divide the dominance. The Galactic ‘residual’ errors dominate the 
spatial scales larger than $\kperp \sim 0.2$, while the extragalactic ‘residual’ errors dominate 
the remaining $\kperp$ modes.

\subsection{Blending impact on the EoR PS and blending ratio tolerance}
\label{res:ResImpactEoR}
\begin{figure*}
  \centering
  \includegraphics[width=0.8\textwidth]{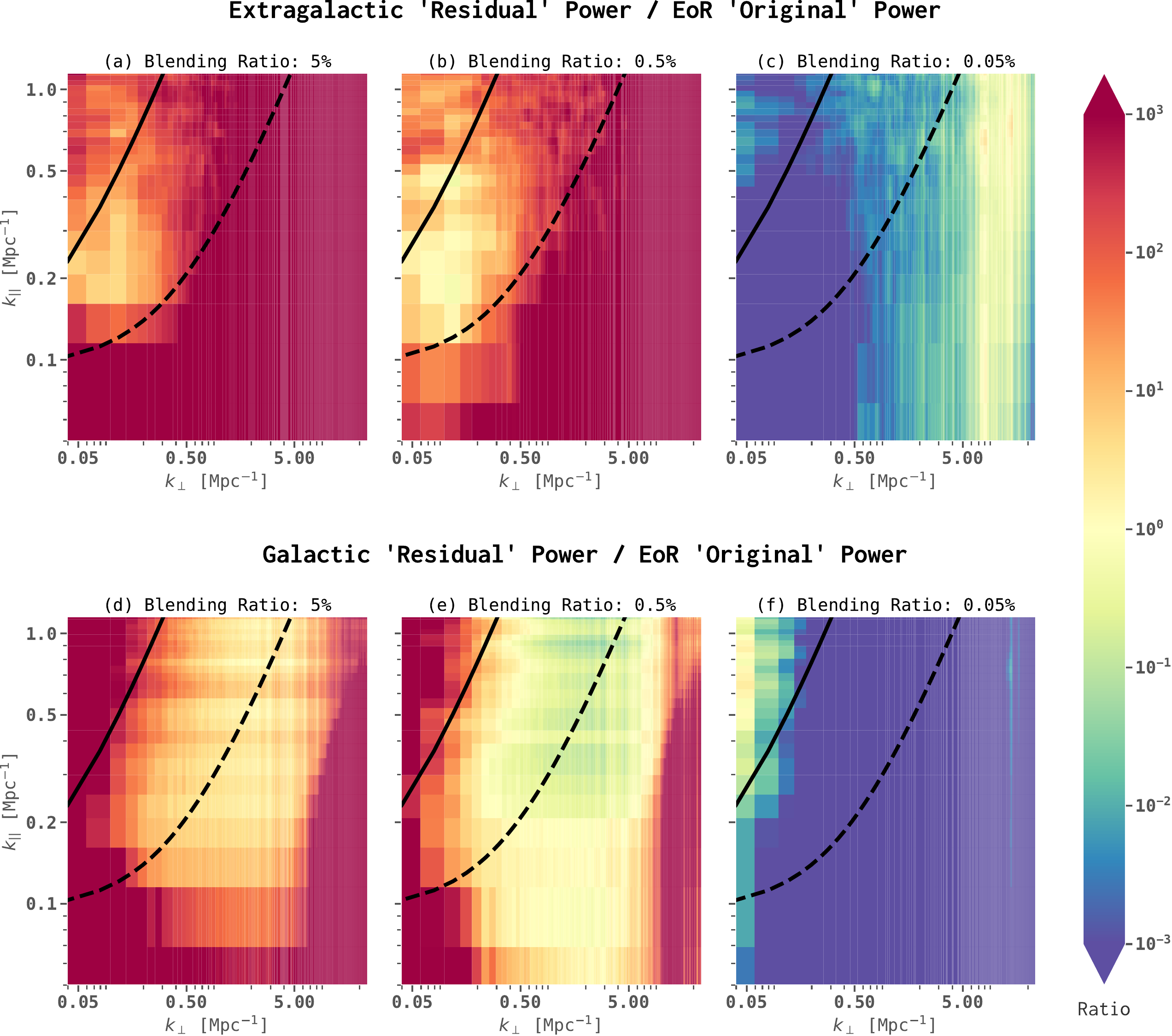}
  \caption{\label{fig:res:ResRatioEoR}%
    The 154 -- 162 \si{\MHz} 2D PS ratios between foreground ‘residual’ powers and EoR ‘original’ powers $R_{\rm res,fg/ori,eor}(\kperp, \klos)$. (Top) $R_{\rm res,ext/ori,eor}(\kperp, \klos)$ ratios under (a) ‘high’ (5 per cent), (b) ‘moderate’ (0.5 per cent), and (c) ‘mild’ (0.05 per cent) blending scenarios. (Bottom) Corresponding $R_{\rm res,gal/ori,eor}(\kperp, \klos)$ ratios under (d) ‘high’ (5 per cent), (e) ‘moderate’ (0.5 per cent), and (f) ‘mild’ (0.05 per cent) blending scenarios.
    Setting $w$ as 2, the black lines define the EoR windows with the dashed and solid lines marking the ‘FoV’ and the ‘horizon’ boundary, respectively.
    Figs. \labelcref{fig:res:RespowerRatio,fig:res:RespowerEvG,fig:res:ResRatioEoR,%
    fig:res:RespowerEvE,fig:res:weighting1,fig:res:weighting2} share the same logarithmic colour bar, showing only six orders of magnitude for easy viewing.
  }
\end{figure*}

The direct quantification of the blending-induced ‘residual’ power impact on EoR detection is achieved by 
calculating the 2D PS ratio between EoR ‘residual’ powers and ‘original’ powers. 
We can infer the specific impact by analysing the contamination within 
the estimated foreground-free EoR window. We plot the $R_{\rm res,fg/ori,eor}(\kperp, \klos)$ 
ratio for the two EoR foregrounds for the three blending cases. From the plotted 
$R_{\rm res,gal/ori,eor}(\kperp, \klos)$ in Fig. \ref{fig:res:ResRatioEoR}, the 
Galactic ‘residual’ powers dominate the large spatial scales, polluting the most important 
EoR $\kperp$ modes significantly. For both the ‘high’ (5 per cent) and ‘moderate’ (0.5 per cent) 
blending ratio cases, 
strong Galactic ‘residual’ powers, which are at least 1000 times the corresponding 
EoR powers, cover almost all the modes within the ‘horizon’-bound region. 
Unanticipatedly, Galactic ‘residual’ powers also contaminate the EoR powers at the 
large \kperp end with a wedge-shaped region at $\klos \gtrsim 7$ and $\klos \gtrsim 9$ 
for the ‘high’ and ‘moderate’ cases, respectively. Recall that the simulated Galactic 
diffuse foregrounds include small-scale fluctuations down to the SKA-Low's angular resolution, 
which enables PS analysis to infer the Galactic impact below the arcmin scale for 
the first time. Although the extragalactic ‘residual’ powers dominate at intermediate 
and small spatial scales, their impact within the EoR window remains significant viewing from 
the plotted $R_{\rm res,ext/ori,eor}(\kperp, \klos)$ in Fig. \ref{fig:res:ResRatioEoR}. 
For the ‘high’ blending ratio case, all of the modes within the ‘FoV’-bound 
region are not recoverable under the impact of either of the foreground components. 
Interestingly, there is a small window of opportunity for the ‘moderate’ case since there 
are modes, which is located at $0.6 \lesssim \kperp \lesssim 5.0$ and 
$0.2 \lesssim \klos \lesssim 1.0$, relatively free of Galactic ‘residual’ power. However, 
this small area vanishes after combining the Galactic and extragalactic foregrounds. For 
the ‘mild’ (0.05 per cent) blending ratio case, significant less ‘residual’ powers 
are present in the measurement space. For one thing, only the largest spatial scales, 
which are located at $0.04 \lesssim \kperp \lesssim 0.08$ and $0.3 \lesssim \klos \lesssim 1.0$, 
are affected by the Galactic ‘residual’ power, leaving most of the ‘FoV’-bound region relatively 
free of impact. For another, the extragalactic ‘residual’ power leaves all ‘FoV’-bound regions 
free of contamination by polluting only the smaller spatial scales outside the EoR window. 
Although caution should be exercised when dealing with Galactic ‘residual’ pollution, 
the impact of ‘residual’ powers on EoR detection is insignificant.

The HEVAL pipeline is designed to isolate source blending as the sole origin of 
errors to evaluate its impact on SKA EoR experiments under a sky-based calibration scheme. 
In the image space, the implementation of \textsc{BlendSim} simulates defects that can be 
attributed solely to source blending. In the visibility space, our analytical modules employ a relative 
calibration scheme to further segregate the calibration errors of blending-induced origin. During the 
map-making phase, wideband, multi-frequency, and spectral fitting procedures mitigate possible 
spectral artefacts that may cloud the impact evaluation in the measurement space. Given these 
measures of impact isolation, we can confidently conclude that the inference in the 2D PS space presented 
in this section originates from blending defects and propagates along the entire EoR analysis pipeline. 
Consider that the ‘residual’ powers are propagated from ‘residual’ visibility-biases that are present 
even after a perfect foreground subtraction, we can draw a blending ratio tolerance to guide the sky-model 
construction for the calibration of SKA EoR experiments. While caution should be taken when addressing Galactic ‘residual’ pollution, the blending ratio tolerance 
for the SKA EoR experiment can be drawn at or slightly below the 0.05 per cent level of the ‘mildly’-blended 
case, which translates to containing only 5 pairs of blended sources per 10000 sources for sky models aimed at 
calibration.

\section{Discussion}
\label{sec:dis}

\subsection{Limitations of the study}
\label{dis:limitation}

Although this paper offers the first systematic insight into source blending in 
the upcoming SKA era, there are limitations to our study.
\subparagraph*{Noise consideration} The simulation of the source-blending effect 
should be considered as the best-case scenario, as the noise impact of the sky-model 
construction is not considered. Given that noise plays a vital 
role during the detection phase of sky-model construction (i.e. the noise level 
determines the detection criteria), a mixture of components often results in a 
measurement error in the total flux density. Considering the accuracy of flux 
density measurements as its own source of sky-model defects, our implantation of 
source-blending effects only considers the flux density reallocation without 
affecting the total flux density of the blended source.  Hence, the total flux 
density of the ‘\textit{ideal}’ and ‘\textit{blended}’ from the same pair remains 
the same.
\subparagraph*{Effects Decoupling} One of the key considerations of this study 
is to infer the calibration errors solely from source-blending defects, 
given that sky-model defects often couple with each other, instrumental 
effects, or a combination of both in practical calibration pipelines. Hence, the 
realization of the HEVAL pipeline in this study decouples source-blending defects 
from other factors, such as polarimetric and side-lobe impact. In particular, 
source-blending defects from sources in the side lobes may introduce additional 
calibration-induced errors in the 2D PS space \citep[see][for a discussion on 
side-lobe effects in the 2D PS space]{2019ApJLi}, which may contaminate different 
parts of the EoR window. We will consider these topics in future studies because of 
the high computational expenses required to evaluate these coupled effects.
\subparagraph*{Temporal effects} The main aim of this study is to infer the frequency-dependent 
gain error. The temporal effects of the complex gain are its own topic 
\citep[see][for detailed discussions on  time-correlated gain effects]{2020MNRASKumar,2022MNRASKumar}. 
We omit consideration of the temporal effect of the complex gain while considering the calibration 
process. Therefore, our results should be viewed as a time-averaged scenario.    
\subparagraph*{Small-scale simulations} The low-frequency radio data used in this study is simulated 
to include physical scales down to the estimated SKA1-Low spatial resolution. Although a commonly 
used practice, which effectively redistributes powers to the simulated small scales, is applied to 
add small-scale fluctuations to our Galactic components, Galactic observations to constrain the 
simulation at such scales are still lacking. For extragalactic foregrounds, our implemented Gaussian 
models of the EDRS populations indicate that fine structures below 6 arcsec are excluded from this study. 
We consider the fine structures of the extended sources as their own topic (i.e. the fidelity of the 
extended source model) and will discuss the impact in an upcoming project. Interested readers can refer 
to \citet{2017PASAProcopio} for a discussion of foreground spatial fidelity impact on EoR foreground subtraction. 
As for the small-scale fluctuations of the EoR simulation, the accuracy of our data is subject to the 
limitations of the semi-analytic models used in \textsc{21cmFAST}.

\subsection{Impact of EoR ‘residual’ powers}
\label{dis:eor}
\begin{figure*}
  \centering
  \includegraphics[width=0.8\textwidth]{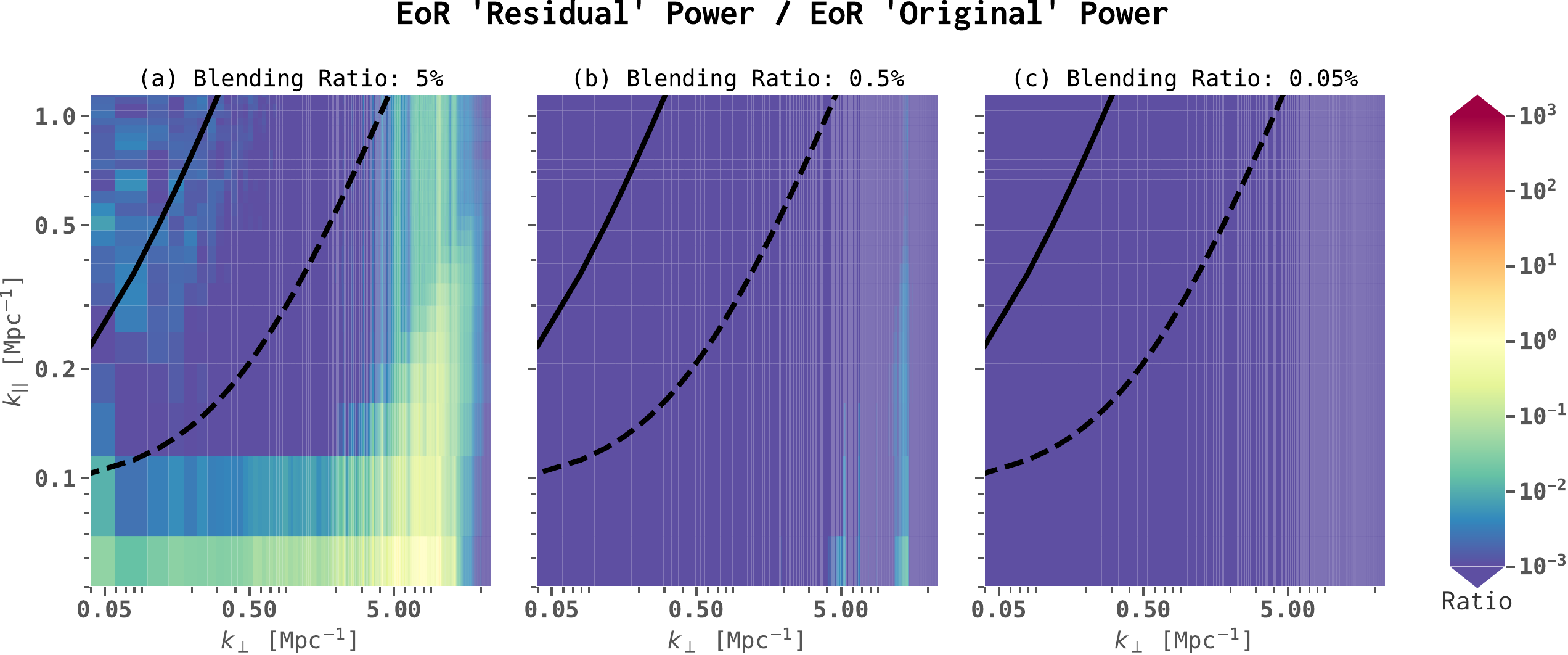}
  \caption{\label{fig:res:RespowerEvE}%
  	The 154 -- 162 \si{\MHz} 2D PS ratios $R_{\rm res\_eor/ori\_eor}(\kperp, \klos)$ (between EoR ‘residual’ and ‘original’ powers) under (a) ‘high’ (5 per cent), (b) ‘moderate’ (0.5 per cent), and (c) ‘mild’ (0.05 per cent) blending scenarios.
    Setting $w$ as 2, the black lines define the EoR windows with the dashed and solid lines marking the ‘FoV’ and the ‘horizon’ boundary, respectively.
    Figs. \labelcref{fig:res:RespowerRatio,fig:res:RespowerEvG,fig:res:ResRatioEoR,%
    fig:res:RespowerEvE,fig:res:weighting1,fig:res:weighting2} share the same logarithmic colour bar, showing only six orders of magnitude for easy viewing.
  }
\end{figure*}

Similar to the calculation of the 2D PS ratio between the foreground ‘residual’ and the EoR 
signal, we can plot the EoR ‘residual’ and ‘original’ power ratio $R_{\rm res,eor/ori,eor}(\kperp, \klos)$ 
and infer the impact of EoR ‘residual’ powers on the EoR signal itself. As demonstrated in Fig. 
\ref{fig:res:RespowerEvE}, only the ‘high’ (5 per cent) blending ratio case has introduced small excess ‘residual’ power within the EoR window. 
However, these powers are expected to have little impact on the EoR detection because they are smaller than one-tenth of the EoR signal and can be considered to be below the noise level. 
The only strong impact is outside the EoR window down to smaller scales, translating to little effect under the foreground avoidance strategy. 
Therefore, unlike the foreground ‘residual’ powers, the propagated EoR ‘residual’ powers originating 
from the ‘highly’-corrupted sky model have little to no direct impact on the EoR detection. However, there might be impact on the accuracy of the EoR parameter inference (such as parameters related to the non-Gaussianity), depending on how precise the model can be constrained under the influence of the EoR ‘residual’ powers. 

\subsection{Impact of imaging weighting on calibration errors}
\label{res:weighting}
\begin{figure*}
  \centering
  \includegraphics[width=1.0\textwidth]{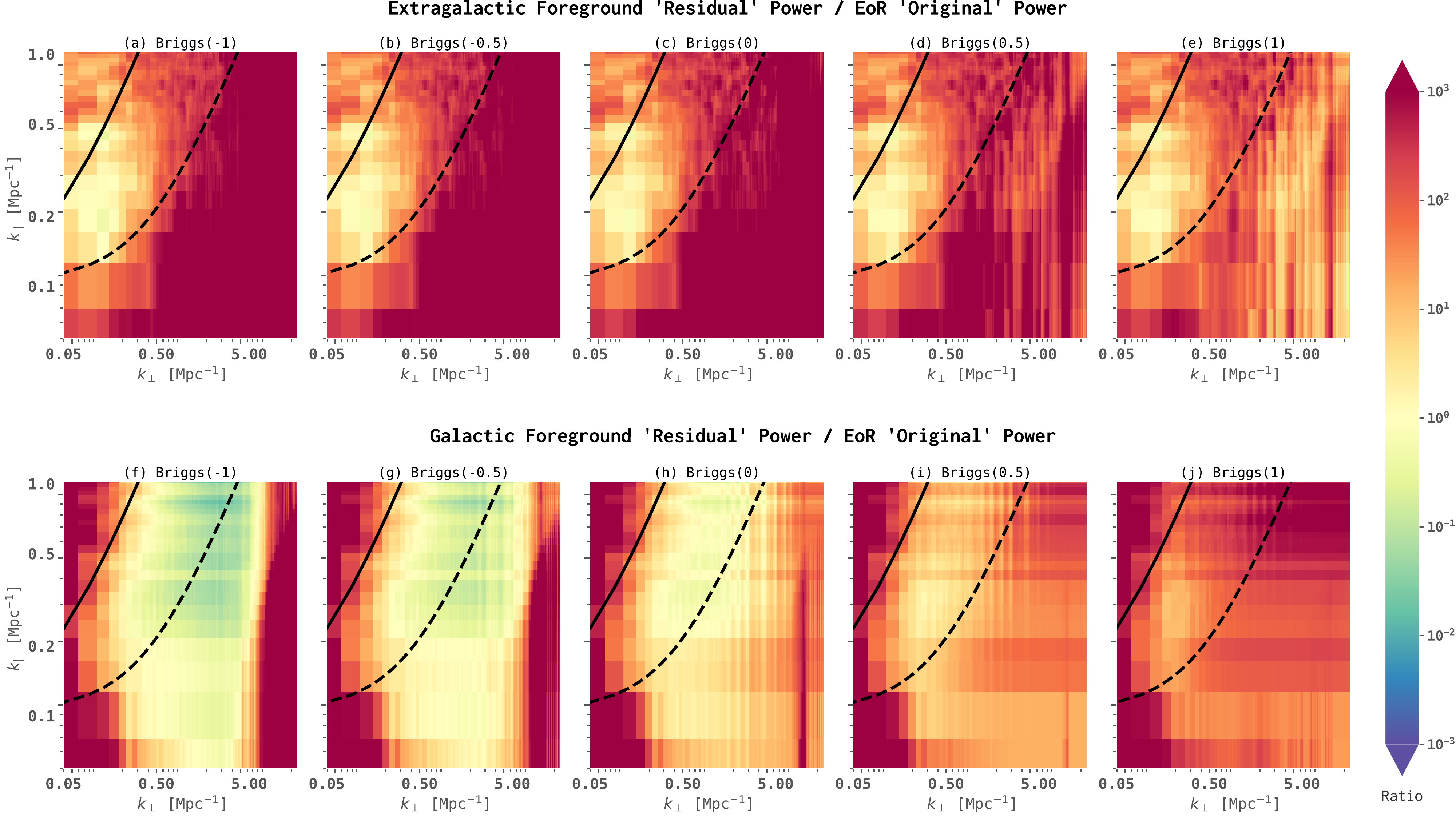}
  \caption{\label{fig:res:weighting1}%
    The 154 -- 162 \si{\MHz} 2D PS ratios between foreground ‘residual’ powers and EoR ‘original’ powers $R_{\rm res,fg/ori,eor}(\kperp, \klos)$ under the ‘moderate’ (0.5 per cent) blending scenario using different Briggs weighting options. (Top) $R_{\rm res,ext/ori,eor}(\kperp, \klos)$ ratios using (a) -1, (b) -0.5, (c) 0, (d) 0.5, and (e) 1 robustness options. (Bottom) Corresponding $R_{\rm res,gal/ori,eor}(\kperp, \klos)$ ratios using (f) -1, (g) -0.5, (h) 0, (i) 0.5, and (j) 1 robustness options.
    Setting $w$ as 2, the black lines define the EoR windows with the dashed and solid lines marking the ‘FoV’ and the ‘horizon’ boundary, respectively.
    Figs. \labelcref{fig:res:RespowerRatio,fig:res:RespowerEvG,fig:res:ResRatioEoR,%
    fig:res:RespowerEvE,fig:res:weighting1,fig:res:weighting2} share the same logarithmic colour bar, showing only six orders of magnitude for easy viewing.
  }
\end{figure*}

\begin{figure*}
  \centering
  \includegraphics[width=1.0\textwidth]{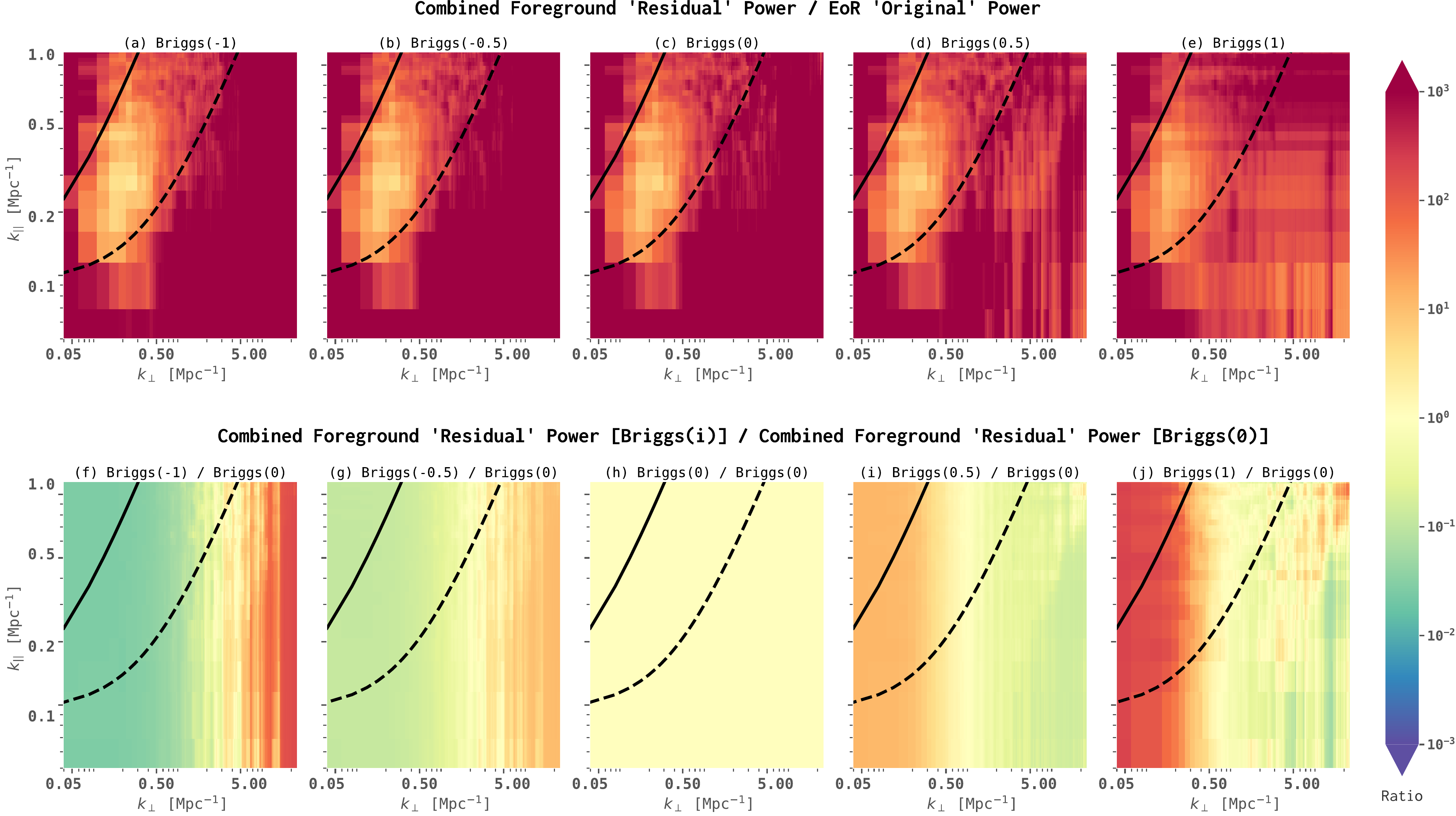}
  \caption{\label{fig:res:weighting2}%
    (Top) The 154 -- 162 \si{\MHz} 2D PS ratios between combined foreground ‘residual’ powers and EoR ‘original’ powers $R_{\rm res,ext+gal/ori,eor}(\kperp, \klos)$ under the ‘moderate’ (0.5 per cent) blending scenario using (a) -1, (b) -0.5, (c) 0, (d) 0.5, and (e) 1 Briggs robustness options, respectively. (Bottom) The 154 -- 162 \si{\MHz} 2D PS ratios between combined foreground ‘residual’ powers using various Briggs robustness options and the Briggs robustness 0 option $R_{\rm res,Briggs(\mathit{i})/res,Briggs(0)}(\kperp, \klos)$ under the ‘moderate’ (0.5 per cent) blending scenario. The various robustness options include (f) -1, (g) -0.5, (h) 0, (i) 0.5, and (j) 1.
    Setting $w$ as 2, the black lines define the EoR windows with the dashed and solid lines marking the ‘FoV’ and the ‘horizon’ boundary, respectively.
    Figs. \labelcref{fig:res:RespowerRatio,fig:res:RespowerEvG,fig:res:ResRatioEoR,%
    fig:res:RespowerEvE,fig:res:weighting1,fig:res:weighting2} share the same logarithmic colour bar, showing only six orders of magnitude for easy viewing.
  }
\end{figure*}

Map-making is a critical step in the ‘reconstructed’-sky PS analysis pipeline. Even with the advanced 
baseline coverage of the upcoming SKA1-Low, the interferometer can only discretely sample the sky 
without unlimited baselines. Thus, image weights are required to determine the filling of the telescope's 
sampling gap during the map-making phase. We discuss the impact of imaging-weight choice on the propagation 
effects of blending-induced calibration errors. We consider the ‘moderate’ (0.5 per cent) blending case, 
which is above the blending ratio tolerance, and add Briggs-weighted clean maps with 
different robustness, including -1, 0, 0.5, and 1, in addition to the -0.5 robustness used in \Cref{subsec:CLEAN}. 
We retain the remaining imaging settings and change only the weight. Thus, we have 5 differently weighted 
maps for the ‘residual’ and ‘original’ image cubes of each foreground component. Fig. \ref{fig:res:weighting1} 
plots the 2D PS ratio between foreground ‘residual’ powers and EoR ‘original’ powers across the 
5 different weighting options. There are clear trends across the different imaging options: (i) there is a reduction in 
‘residual’ power at the smallest spatial scales with robustness varying from -1 to 1; (ii) the Galactic ‘residual’ power 
at the intermediate spatial scales increases with robustness varying from -1 to 1; and (iii) the ‘residual’ power at the 
largest spatial scales increases as the robustness varies from -1 to 1. 
Given these trends, we can identify a potential mitigation strategy for blending-induced ‘residual’ powers by 
choosing the suitable imaging weight for the particular observing sky during the map-making phase. As shown in Fig. 
\ref{fig:res:weighting2}, a viable approach is using the robustness -1 option for this particular simulated sky patch 
considering the reduced impact in the measurement space. 
There is an evident boost in the detection possibility at the largest spatial scales, which can be inferred from the 
weighting ratio figures in the second row of Fig. \ref{fig:res:weighting2}. From a general EoR detection point-of-view, 
a more holistic approach should be considered by balancing between different sources of errors across the analysis pipeline 
with each individual bias factor identified, evaluated, and mitigated.

\subsection{Mitigation of blending-induced effects}
\label{dis:mitigation}

To reduce the propagated ‘residual’ power discussed in this paper, upcoming SKA EoR experiments require 
dedicated de-blending efforts built directly within the sky-model construction pipeline. Current de-blending 
efforts based on LOFAR observations heavily involve human input \citep[see][for the LOFAR de-blending workflow]{2019A&AWilliams,2021A&AKondapally}. 
With the upcoming SKA1-Low's improved sensitivity compared to the current observations, source blending 
could only be exacerbated. Therefore, automatic de-blending approaches, such as those applied 
in the optical domain \citep[see][for a review]{2021NatRPMelchior}, are required to address the blending 
issue both in decomposition and association, but the implementation is still an open 
question. To aid the development of de-blending pipelines and strategies, the developer 
community needs proper datasets through observations and simulations. From an 
observational perspective, obtaining higher-resolution radio observations 
of the SKA EoR fields through the international LOFAR Telescope \citep{2011Vermeulen,2016A&AJackson} would both 
offer a direct blending effect deduction and help with the development of de-blending methods during decomposition. 
Multi-frequency synergies, such as joint observations with 
\textit{JWST}\footnote{%
  \url{https://www.jwst.nasa.gov/}} \citep{2023PASPMcElwain}, 
the Large Synoptic Survey Telescope (LSST\footnote{%
  \url{https://www.lsst.org/}}; \citealt{2019ApJIvezic}), 
or the Chinese Survey Space Telescope (CSST\footnote{%
  \url{http://www.bao.ac.cn/csst/}}; \citealt{2011SSPMAZhan}), of the SKA observing sky, would also be crucial 
for the SKA to battle source blending during the source-association phase. From a simulation perspective, a 
high-fidelity multi-frequency (across the radio to infrared and optical bands) EDRS simulation is required to 
develop de-blending methods. 
An alternative approach to direct de-blending efforts is to treat blending-induced ‘residual’ errors in a similar 
manner to other frequency-dependent errors. General frequency-dependent error-mitigation strategies include the 
usage of spectrally-smoothed antennas \citep{2016MNRASBarry}, baseline weighting \citep{2017MNRASEwall-Wice}, and 
simultaneous multi-channel calibration \citep{2015MNRASYatawatta}. In practice, a combination 
of both a direct de-blending pipeline and indirect error-mitigation strategies might be used to achieve an optimal 
solution that effectively attenuates the blending-induced impact.

\section{Conclusion}
\label{sec:con}
We make the first attempt to systematically assess the impact of source blending 
in the low-frequency radio band for the upcoming SKA1-Low. This study introduces 
a clear definition of source blending in the radio window, identifies the two-fold blending 
defect in both the spatial and frequency domains, and investigates the underlying impact 
of source blending on interferometric calibrations for SKA EoR experiments. We 
summarize our work as follows:
\begin{enumerate}
\item Using the quick-and-dirty method presented in \Cref{sec:blending}, which utilizes the 
W08 source model and the latest SKA1-Low array configuration, we estimate an extended 
level of blending ($\sim 5 - 28$ per cent sources are blended) for the upcoming SKA1-Low 
observing sky and identified source blending as one of the key imperfections impacting the 
sky-model construction for SKA EoR experiments. 
\item By using our HEVAL pipeline (which simulates the low-frequency radio sky containing physical 
scales spatially resolvable by the SKA1-Low) to quantify the blending-induced calibration impact 
in a sky-based scheme for EoR detections, we find that frequency-dependent calibration errors 
from poor calibration against blending-corrupted sky models coupled with strong foregrounds leave 
additive ‘residual’ powers within the EoR window in the 2D PS space and may significantly 
impede EoR detections. Our findings corroborate with both the B16 simulation approach and the E17 
analytic analysis results, albeit focusing on three different sky-model defects. Furthermore, 
by considering the extragalactic and Galactic foregrounds separately, we conclude that the extragalactic 
and Galactic ‘residual’ powers contaminate EoR scales at larger \kperp and smaller \kperp, respectively, 
with a clear division around $\kperp \sim 0.2$. 
\item To determine the blending tolerance for SKA EoR experiments, we perform three tests with relatively 
low levels of blending ratios (5, 0.5, and 0.05 per cent). The tests show that the blending defects 
at the former two levels introduce strong ‘residual’ powers in the measurement space, seriously contaminating 
the key EoR scales ($0.1 \lesssim k \lesssim 2$ $\si{\Mpc}^{-1}$) within the EoR window, whereas sky models 
with a blending ratio at 0.05 per cent leave little impact in the measurement space. Hence, a blending ratio 
of approximately 0.05 per cent is identified as the blending tolerance for the upcoming SKA1-Low sky-model 
construction.
\item Given that our estimated $5 - 28$ per cent blending ratio for the SKA1-Low has quite a large margin 
with a tolerance of 0.05 per cent, we discuss the possibility of mitigating and suppressing the 
blending-induced impact through different imaging weights. For the designed baseline of the SKA1-Low, 
Briggs weighting with robustness -1 would be a better choice to mitigate the blending-induced impact 
at key EoR scales.
\item To directly mitigate the impact of source-blending defects, de-blending techniques must be involved 
in both the SKA source detection and sky-model construction to pass the required blending tolerance. From 
a software perspective, the source detection community should prioritize including de-blending or blending 
flagging during both the component-decomposition and -association phases. Additionally, priority should 
be set to port and adapt existing automatic de-blending methods, such as those used in the optical bands, to 
the radio bands. From an observation perspective, adding higher-resolution or less-blended observations to 
the SKA EoR fields via synergies with the international LOFAR Telescope and instruments from other wavelength 
bands will be crucial for the SKA to reduce the blending level through cross-match assessments directly. In 
practice, joint efforts from the software and observation perspectives are most likely required to contain 
blending-induced errors within the necessary overall EoR error budget.
\end{enumerate}

\section*{Acknowledgements}
We thank Weitian Li for valuable discussions regarding \textsc{fg21sim}, Andrei Mesinger, Yuxiang Qin, 
and Steven Murray for their help with \textsc{21cmFAST}, Fred Dulwich for technical support in resolving 
issues with \textsc{OSKAR}, and André Offringa for guidance on map-making with \textsc{WSClean}. 
We thank Hekun Lee, Benjamin McKinley, and Junhua Gu for scientific discussions, Sanjay Bhatnagar for 
reading the manuscript and providing insightful comments, and Bingxue Yang's help with the preparation 
of the final draft of this manuscript. This work was performed on the Gravity Cluster of Department of 
Astronomy (DOA) at Shanghai Jiao Tong University (SJTU). Some of the results in this paper have been 
derived using the healpy and HEALPix packages. 
This work is supported by the Ministry of Science and Technology of China (Grant No. 2020SKA0110201), 
and the National Natural Science Foundation of China (Grant Nos. 11973033 and 11835009). 
ZZH acknowledges the support from the National Science Foundation of China (Grant No. 12203085). 
SVW acknowledges the financial assistance of the South African Radio Astronomy Observatory (SARAO; https://www.sarao.ac.za). 
Parts of this research were supported by the Australian Research Council Centre of Excellence for All Sky Astrophysics in 3 Dimensions (ASTRO 3D), through project number CE170100013.

\section*{Data Availability}

The end-to-end simulate suite \textsc{Fg21Sim+} is an open-source project developed at \url{https://github.com/Fg21Sim/}. 
Owing to the large data volume, the data generated by this work will be shared under a reasonable request at 
\url{https://github.com/Fg21Sim/Data}. Datasets with the same specifications as in this study can be simulated 
using \textsc{Fg21Sim+} with the parameters provided by the authors.

\section*{Author Contribution Statement}
Here is the list of contributions:
\begin{itemize}
  \item C.Shan: project proposal, coding (blending ratio estimation and the majority of HEVAL pipeline), methodology, testing, validation, result analysis, and writing (original draft and modification).
  \item H.Xu: scientific feedback, results discussion, writing (feedback and modification), and funding support.
  \item Y.Zhu: coding (contribution to the \textsc{ESim} and \textsc{BlendSim} modules), validation, results analysis (feedback and discussion), and writing (feedback and modification).
  \item Y.Zhao: scientific feedback, results analysis (feedback and discussion), and writing (feedback and modification).
  \item S.White: scientific feedback, figure modification, and writing (feedback and modification).
  \item J.Line: scientific feedback, results analysis (feedback and discussion) and writing (feedback).
  \item D.Zheng: methodology (feedback and discussion) and coding (contribution to the \textsc{GSim}).
  \item Z.Zhu: scientific feedback and writing (feedback).
  \item D.Hu: scientific feedback and writing (feedback).
  \item Z.Zhang: scientific feedback.
  \item X.Wu: scientific and funding support.
\end{itemize}



\bibliographystyle{mnras}
\bibliography{paper_blending_reference_intro,paper_blending_reference_meth,paper_blending_reference_rest} 




\appendix

\section{Small-scale fluctuation of Galactic foregrounds}
\label{app:smallscales}

In this section, we introduce our method of adding small-scale fluctuations 
to the Galactic emission in detail.  
First, \textsc{GSim} extrapolates the angular PS of the basis 
templates to the smaller scales and generates GRFs, $G_{\mathrm{grf}}$, 
based on the extrapolated angular PS. We adopt the angular 
power spectrum model for the synchrotron and free-free emission as
\begin{equation}
\mathcal{C}_{\ell}^{\mathrm{sync}} = \ell^\gamma \left[ 1 - \mathrm{exp}  \left( -\ell^2 \sigma_{\mathrm{temp}}^2 \right) \right] \label{eqn:Gsync-cl}
\end{equation}
\citep{2015MNRASRemazeilles} and
\begin{equation}
\mathcal{C}_{\ell}^{\mathrm{free}} = \ell^\gamma \left[ \mathrm{exp} \left( -\ell^2 \sigma_{\mathrm{sim}}^2 \right) - \mathrm{exp}  \left( -\ell^2 \sigma_{\mathrm{temp}}^2 \right) \right] \label{eqn:Gfree-cl}
\end{equation}
\citep{2013A&ADelabrouille}, respectively, where $\sigma_{\mathrm{temp}}$ 
and $\sigma_{\mathrm{sim}}$ are the full width at half-maximum (FWHM) 
beam of the basis template and simulated sky maps, respectively, and the $\gamma$ 
index is the power law index of the basis template. By adopting the two 
models, the GRF maps ($G_{\mathrm{grf}}$) of the two Galactic diffuse 
components can be simulated using \textsc{GSim} by converting 
$\mathcal{C}_{\ell}$ to maps with angular scales from degree level down 
to the required $\ell$. To compensate for the high demand of both memory 
and CPU for generating high spatial resolution GRFs, we employ the \textsc{GSim} 
module with a custom \textsc{cl2alm} tool modified from the 
\textsc{healpy}\footnote{%
  \textsc{healpy}: \url{https://github.com/healpy/healpy}}  
package \citep{2019JOSSZonca} and \textsc{ducc0}\footnote{%
  \textsc{ducc0}: \url{https://gitlab.mpcdf.mpg.de/mtr/ducc/-/tree/ducc0}} 
with multi-threaded support. The former calculates the spherical harmonic 
coefficients ($a_{lm}$) from the angular PS, and the latter offers 
map realization from the coefficients. Then, the small-scale fluctuation map 
($G_{\mathrm{ss}}$) is generated with a whitened $G_{\mathrm{grf}}$, 
which has zero mean and unit variance, using
\begin{equation}
G_{\mathrm{ss}} = \alpha G_{\mathrm{grf}} \, G_{\mathrm{temp}}^\beta, \label{eqn:Gss}
\end{equation}
where $\alpha$ and $\beta$, which dictate the mean and variance of the small-scale 
maps, respectively, are custom parameters to be determined. Finally, the simulated Galactic 
emission with the addition of small-scale fluctuations ($G_{\mathrm{sim}}$) 
can be inferred with
\begin{subequations}\label{eqn:Gsim}
\begin{align}
G_{\mathrm{sim}} &= G_{\mathrm{temp}} + G_{\mathrm{ss}}, \label{eqn:Gsim-full}\\
& = G_{\mathrm{temp}} + \alpha G_{\mathrm{grf}} \, G_{\mathrm{temp}}^\beta. \label{eqn:Gsim-detail}
\end{align}
\end{subequations}

For each simulation run, \textsc{GSim} offers on-the-fly fitting of the 
three key parameters, $\alpha$, $\beta$, and $\gamma$, to prescribe the 
generation of sky maps. The fitting process is based on the 
\textsc{emcee}\footnote{%
  \textsc{emcee}: \url{https://emcee.readthedocs.io/en/stable/}} \citep{2013PASPForemanMackey} 
Python library of the  Markov Chain Monte Carlo (MCMC) method. To fit the 
power law index $\gamma$, \textsc{GSim} employs the angular PS 
analysis of the Galactic all-sky basis template on a Hierarchical Equal 
Area and isoLatitude Pixelization (HEALPix\footnote{%
  HEALPix: \url{http://healpix.sf.net}}, \citealt{2005ApJGorski}) 
sphere using the \textsc{anafast}\footnote{%
  \textsc{anafast}: \url{https://healpix.sourceforge.io/html/fac_anafast.htm}} module. 
Subsequently, the $\gamma$ index is inferred by fitting a power law to the 
large angular scale ($\ell\lesssim 500$) of the computed $\mathcal{C}_{\ell}$. 
As for the fitting of $\alpha$ and $\beta$, we use a set of statistical constraints, 
which require the preservation of the underlying statistical properties of the 
basis map with the addition of small-scale features (equation \ref{eqn:Gsim-detail}). 
To achieve these constraints, \textsc{GSim} uses a model that combines (i) an equivalence 
test, which is implemented through a two one-sided t-tests (TOST, \citealt{1987Schuirmann}) 
procedure, (ii) a maximum mean discrepancy (MMD, \citealt{1953Fortet}) test, 
and (iii) difference minimization of the mean, variance, skewness, and kurtosis  
between the basis template ($G_{\mathrm{temp}}$) and the feature-added map 
($G_{\mathrm{temp}} + \alpha G_{\mathrm{grf}} \, G_{\mathrm{temp}}^\beta$).

For the purpose of this work, the $\gamma$ index of synchrotron and free-free 
component is fitted as -2.220 and -2.426, respectively, using $\mathcal{C}_{\ell}$ 
with $\ell$ ranging from 30 to 90. However, due to the high spatial resolution 
requirement of the small-scale features (a HEALPix map with Nside = 32,768 
meets the SKA spatial resolving power), we fit $\alpha$ and $\beta$ using partial 
sky coverage instead of the full sky map to avoid the extremely high CPU and memory 
costs. By fitting to the aimed sky coverage (R.A., Dec.\@ = \SI{0}{\degree}, \SI{-27}{\degree}), 
we find the best-fitting result for the synchrotron ($\alpha=0.0342$, $\beta=0.227$) 
and free-free emission ($\alpha=0.00785$, $\beta=0.526$).

\section{Deriving the gain error solution \& per-baseline bias}
\label{app:full-derivation}

This section details our formalism of the analytic analysis to evaluate the blending-induced 
calibration impact. Recall that we have established the sky-based per-frequency 
per-antenna calibration scheme in \Cref{subsubsec:skycali} and presented key equations in 
\Cref{subsubsec:cali-error}. 
Here, we start by deriving the linear systems of equations 
for the per-antenna gain amplitude and phase error in \Cref{app:gain-error} using the PWL logarithmic implementation\footnote{%
  Although we note the drawbacks compared to the linearized approximation \citep[see][for a discussion]{2010MNRASLiu}, the usage of logarithmic implementation makes the derivation of relative residual gain errors simple and computationally cheap. Careful readers may already infer that there is no impact of the error solution on the estimated propagation bias by connecting equations \eqref{eqn:comb-vis-pair} and \eqref{eqn:bias}.}. 
Then, \Cref{app:error-sol} presents the SVD-based method used to infer the least-squares solution. 
Finally, the per-frequency per-baseline propagation visibility bias is estimated in \Cref{app:prop-error}.

\subsection{Deriving the per-antenna gain amplitude and phase error}
\label{app:gain-error}

With our paired sky model approach, we can realize two equations using the measured visibility equation (equation \ref{eqn:visibility-form2}) for applying an ‘\textit{ideal}’ sky model (equation \ref{eqn:visibility-ideal}) and a ‘\textit{blended}’ (equation \ref{eqn:visibility-blend}) sky model, respectively. 
For each cross-correlation $V_{ij}$, we can determine the calibration 
error of the ill-calibrated scenario by combining equations (\ref{eqn:visibility-ideal}) and 
(\ref{eqn:visibility-blend}) and solve
\begin{equation}\label{appeqn:comb-vis}
  \begin{multlined}
    \mathrm{exp} \left[ \left( \eta_i + \eta_j \right) + \mathrm{i} \left( \phi_j -\phi_i \right) \right] V_{ij}^{\mathrm{ideal}} + n_{ij} \\
    = \mathrm{exp} \left[ \left( \acute{\eta}_i + \acute{\eta}_j \right) + \mathrm{i} \left( \acute{\phi}_j -\acute{\phi}_i \right) \right] \acute{V}_{ij}^{\mathrm{blend}} + n_{ij}.
  \end{multlined}
\end{equation}
Here, we effectively eliminated the impact of the baseline noise, since the 
noise item $n_{ij}$ is specific to the baseline $B_{ij}$ under the assumption 
made in equation (\ref{eqn:correlation-c}). By taking the logarithmic form of 
equation (\ref{appeqn:comb-vis}), we have 
\begin{equation}
\left( \eta_i + \eta_j \right) + \mathrm{i} \left( \phi_j -\phi_i \right) + \ln V_{ij}^{\mathrm{ideal}} = \left( \acute{\eta}_i + \acute{\eta}_j \right) + \mathrm{i} \left( \acute{\phi}_j -\acute{\phi}_i \right) + \ln \acute{V}_{ij}^{\mathrm{blend}}.  \label{appeqn:visibility-combine}
\end{equation}

By substituting the gain error 
($\Delta \eta_i \equiv \eta_i - \acute{\eta}_i $ and 
$\Delta \phi_i \equiv \phi_i - \acute{\phi}_i$, respectively) and the visibility 
difference in terms of amplitude and phase 
($R_{ij} \equiv \ln \abs{\acute{V}_{ij}^{\mathrm{blend}}} - \ln \abs{V_{ij}^{\mathrm{ideal}}}$ 
and $I_{ij} \equiv \arg \abs{\acute{V}_{ij}^{\mathrm{blend}}} - \arg \abs{V_{ij}^{\mathrm{ideal}}}$, 
respectively), 
equation \eqref{appeqn:visibility-combine} can be decoupled as two sets of linear 
equations presented in equation \eqref{eqn:comb-vis-pair}.

\subsection{Gain error solution}
\label{app:error-sol}

To solve the two sets of linear systems of equations (equation \ref{eqn:comb-vis-pair}), we proceed by 
rewriting equation \eqref{eqn:comb-vis-real} and equation \eqref{eqn:comb-vis-imag} in two matrix systems. 
Taking equation \eqref{eqn:comb-vis-real} as an example, we have
\begin{subequations}\label{appeqn:real-matrix-pair}
\begin{align}
\underbrace{
\begin{bmatrix} 
& 1 & 1 & 0 & \dots  & 0 & 0 & 0 &\\
& 1 & 0 & 1 & \dots  & 0 & 0 & 0 &\\
& 1 & 0 & 0 & \dots  & 0 & 0 & 0 &\\
& \vdots & \vdots & \vdots & \ddots & \vdots & \vdots & \vdots &\\
& 1 & 0 & 0 & \dots  & 0 & 0 & 1 &\\
& \vdots & \vdots & \vdots & \ddots & \vdots & \vdots & \vdots &\\
& 0 & 0 & 0 & \dots  & 1 & 1 & 0 &\\
& 0 & 0 & 0 & \dots  & 1 & 0 & 1 &\\
& 0 & 0 & 0 & \dots  & 0 & 1 & 1 &
\end{bmatrix}}_{\displaystyle \equiv \bm{\mathsf{A}}_R}
\underbrace{
\begin{bmatrix}
\Delta \eta_1 \\
\Delta \eta_2 \\
\Delta \eta_3 \\
\vdots \\
\Delta \eta_N
\end{bmatrix}}_{\displaystyle \equiv \bm{x}_R}
=
\underbrace{
\begin{bmatrix}
R_{12} \\
R_{13} \\
R_{14} \\
\vdots \\
R_{1N}\\
\vdots \\
R_{N-2N-1}\\
R_{N-2N}\\
R_{N-1N}
\end{bmatrix}}_{\displaystyle \equiv \bm{b}_R}
, \label{appeqn:real-matrix-full}\\
\underset{\bm{\tilde{x}}_R}{\mathrm{argmin}} \norm{\bm{\mathsf{A}}_R \bm{\tilde{x}}_R - \bm{b}_R}_2, \label{appeqn:real-matrix-sol}
\end{align}
\end{subequations}
where $\bm{\mathsf{A}}_R$, which will be referred to as the array configuration 
matrix, is a $N \times C^2_N$ matrix, $\bm{x}_R$ is a $N \times 1$ matrix, 
and $\bm{b}_R$ is a $C^2_N \times 1$ matrix. In general, there will 
not be enough degrees of freedom in $\bm{x}_R$ when solving $N$ unknown 
$\Delta \eta$ with $C^2_N$ measurements. Thus, the least-squares solution 
is solved and satisfies the optimization in equation \eqref{appeqn:real-matrix-sol}.
In a similar fashion, we can also write the set of the phase difference 
equations in a matrix form and solve the least-squares solution:
\begin{subequations}\label{appeqn:imag-matrix-pair}
\begin{align}
\bm{\mathsf{A}}_I \bm{x}_I = \bm{b}_I, \label{appeqn:imag-matrix-full}\\
\underset{\bm{\tilde{x}}_I}{\mathrm{argmin}} \norm{\bm{\mathsf{A}}_I \bm{\tilde{x}}_I - \bm{b}_I}_2. \label{appeqn:imag-matrix-sol}
\end{align}
\end{subequations}

To solve those overdetermined systems of equations, we use the SVD 
\citep[see][for a review]{1993Stewart} to perform the pseudoinverse 
of the array configuration matrix and find the least-squares solution 
\citep[see][for detailed demonstrations]{2019brunton_kutz}. 
According to the Moore-Penrose left pseudoinverse 
\citep{1955Penrose,1956Penrose,1965Rohde,1970Zlobec}, $\bm{\mathsf{A}}$ matrix 
(referring to for both the amplitude and phase array configuration matrix) has a 
pseudoinverse $\bm{\mathsf{A}}^\dagger$, and the least-squares solution of 
the blending-induced amplitude and phase error, 
$\bm{\tilde{x}}_R$ and $\bm{\tilde{x}}_I$, can be solved via
\begin{equation}
\bm{\tilde{x}} = \bm{\mathsf{A}}^\dagger \bm{b}. \label{appeqn:minimize-real}
\end{equation}

\subsection{Deriving the per-baseline propagation error}
\label{app:prop-error}

The aftermath of the calibration against a sky model with source-blending 
defects is further discussed here. As mentioned in \Cref{subsubsec:skycali}, 
after solving each antenna gain factor, $g_i$, the calibration solution 
will be applied to the observed visibility to recover the true cross-correlation of 
the sky signal. In the ill-calibrated scenario, the blending-induced antenna 
gain errors, $\Delta \eta$s and $\Delta \phi$s, will propagate to the 
calibrated visibilities of the sky signal for each baseline and leave a 
per-frequency per-baseline propagation bias within the visibility space. 

Using the least-squares solution of both the amplitude and phase error 
matrix ($\bm{x}_R$ and $\bm{x}_I$, respectively) of the blending-induced 
antenna gain, we can further infer the per-frequency per-baseline propagation 
bias upon the visibilities of the sky signal. To infer the poor-calibration 
impact on the SKA EoR experiments, a set of simulated data with instrumental responses 
and resolution fidelity, including 
extragalactic foregrounds, Galactic foregrounds, and the 21-cm signal, are utilized to estimate 
the propagation bias for each baseline in a realistic fashion. Here, taking Galactic foregrounds as an 
example, we can rewrite the measured visibility equation (equation \ref{eqn:visibility-form2}) 
and its logarithmic form using the true visibility $S^{\mathrm{gal-true}}_{ij}$ 
and ill-calibrated visibility $\acute{S}^{\mathrm{gal-ill}}_{ij}$ of Galactic 
emission as
\begin{equation}\label{appeqn:comb-sky}
  \begin{multlined}
    \mathrm{exp} \left[ \left( \eta_i + \eta_j \right) + \mathrm{i} \left( \phi_j -\phi_i \right) \right] S_{ij}^{\mathrm{gal-true}} + n_{ij}\\
    = \mathrm{exp} \left[ \left( \acute{\eta}_i + \acute{\eta}_j \right) + \mathrm{i} \left( \acute{\phi}_j -\acute{\phi}_i \right) \right] \acute{S}_{ij}^{\mathrm{gal-ill}} + n_{ij}
  \end{multlined}
\end{equation}
and
\begin{equation}
\left( \eta_i + \eta_j \right) + \mathrm{i} \left( \phi_j -\phi_i \right) + \ln S_{ij}^{\mathrm{gal-true}} = \left( \acute{\eta}_i + \acute{\eta}_j \right) + \mathrm{i} \left( \acute{\phi}_j -\acute{\phi}_i \right) + \ln \acute{S}_{ij}^{\mathrm{gal-ill}}, \label{appeqn:sky-log}
\end{equation}
respectively. Similar to the decoupling of equation \eqref{appeqn:visibility-combine}, 
we further decompose equation \eqref{appeqn:sky-log} into the real and imaginary parts 
separately and substitute the calibrated gain errors. Therefore, 
equation \eqref{appeqn:sky-log} takes the form of
\begin{subequations}\label{appeqn:diff-sky-pair}
\begin{align}
\ln \abs{\acute{S}_{ij}^{\mathrm{gal-ill}}} - \ln \abs{S_{ij}^{\mathrm{gal-true}}} &= \Delta \eta_i + \Delta \eta_j, \label{appeqn:diff-sky-real}\\
\arg \abs{\acute{S}_{ij}^{\mathrm{gal-ill}}} - \arg \abs{S_{ij}^{\mathrm{gal-true}}} &= \Delta \phi_j - \Delta \phi_i. \label{appeqn:diff-sky-imag}
\end{align}
\end{subequations}
Since all the $\Delta \eta$s and $\Delta \phi$s were solved for each frequency individually 
using methods described in \Cref{app:error-sol}, the per-frequency per-baseline 
propagation amplitude and phase bias of Galactic emission can be inferred 
using equation \eqref{appeqn:diff-sky-pair} by taking the exponential form of equation \eqref{appeqn:diff-sky-real}:
\begin{subequations}\label{appeqn:bias-galactic}
\begin{align}
\abs{\acute{S}_{ij}^{\mathrm{gal-ill}}} - \abs{S_{ij}^{\mathrm{gal-true}}} &= \left[ \mathrm{exp} \left( \Delta \eta_i + \Delta \eta_j \right) - 1 \right] \abs{S_{ij}^{\mathrm{gal-true}}} , \label{appeqn:bias-sky-real-galactic}\\
\arg \abs{\acute{S}_{ij}^{\mathrm{gal-ill}}} - \arg \abs{S_{ij}^{\mathrm{gal-true}}} &= \Delta \phi_j - \Delta \phi_i. \label{appeqn:bias-sky-imag-galactic}
\end{align}
\end{subequations}
Similarly, we also derive the amplitude and phase bias for the other components. 
Hence, the per-frequency per-baseline propagation bias of all the sky components in the 
visibility space can be estimated in the general form, as shown in equation \eqref{eqn:bias}.

\begin{table*}
 \caption{Calibration-related mathematical notions used in the paper.}
 \label{tab:allsymbols}
 \begin{tabular}{ll}
  \hline
  Notion & Meaning \\
  \hline
  $N$ & Number of antennas \\[2pt]
  $f$ & Frequency channel \\[2pt]
  $\Delta t$ & Observation time interval \\[2pt]  
  $v_i$ & Measured voltage of the $i$th antenna \\[2pt]
  $v^*_i$ & Complex conjugate of the antenna measured voltage $v_i$ \\[2pt]
  $s_i$ & True sky signal received by the $i$th antenna \\[2pt]
  $s^*_i$ & Complex conjugate of the true sky signal $s_i$ \\[2pt]
  $g_i$ & Complex gain of the $i$th antenna \\[2pt]
  $g^*_i$ & Complex conjugate of the antenna complex gain $g_i$ \\[2pt]
  $\acute{g}_i$ & Ill-calibrated gain of the $i$th antenna due to the blending effect \\[2pt]
  $\eta_i$ & Amplitude of the antenna gain $g_i$ \\[2pt]
  $\acute{\eta}_i$ & Ill-calibrated antenna gain amplitude of the $i$th antenna due to the blending effect \\[2pt]
  $\Delta \eta_i$ & Amplitude difference of the blending-induced per-antenna gain error: difference between $\eta_i$ and $\acute{\eta}_i$ \\[2pt]
  $\phi_i$ & Phase of the antenna gain $g_i$ \\[2pt]
  $\acute{\phi}_i$ & Ill-calibrated antenna gain phase of the $i$th antenna due to the blending effect \\[2pt]
  $\Delta \phi_i$ & Phase difference of the blending-induced per-antenna gain error: difference between $\phi_i$ and $\acute{\phi}_i$ \\[2pt]
  $n_i$ & Noise of the ith antenna \\[2pt]
  $n^*_i$ & Complex conjugate of the antenna noise $n_i$ \\[2pt]
  $n_{ij}$ & Noise of the $\bm{{B}}_{ij}$ baseline \\[2pt]
  $\bm{{B}}_{ij}$ & Baseline vector of the $i$th and $j$th antenna \\[2pt]
  $V_{ij}$ & Visibility correlated by baseline $\bm{{B}}_{ij}$ \\[2pt]
  $V_{ij}^{\mathrm{ideal}}$ & Visibility of an ‘\textit{ideal}’ sky model correlated by the baseline $\bm{{B}}_{ij}$ \\[2pt]
  $\acute{V}_{ij}^{\mathrm{blend}}$ & Visibility of a ‘\textit{blended}’ sky model correlated by the baseline $\bm{{B}}_{ij}$ \\[2pt]
  $S_{ij}$ & True sky signal visibility correlated by baseline $\bm{{B}}_{ij}$ \\[2pt]
  $S^{\mathrm{true}}_{ij}$ & True visibility of each sky component correlated by baseline $\bm{{B}}_{ij}$ \\[2pt]
  $\acute{S}_{ij}^{\mathrm{ill}}$ & Ill-calibrated visibility of each sky component correlated by baseline $\bm{{B}}_{ij}$ \\[2pt]
  $G_{ij}$ & Gain factor of the $\bm{{B}}_{ij}$ baseline \\[2pt]
  $R_{ij}$ & Logarithmic amplitude difference between the ‘\textit{ideal}’ and ‘\textit{blended}’ sky model visibility correlated by the baseline $\bm{{B}}_{ij}$ \\[2pt]
  $I_{ij}$ & Logarithmic phase difference between the ‘\textit{ideal}’ and ‘\textit{blended}’ sky model visibility correlated by the baseline $\bm{{B}}_{ij}$ \\[2pt]
  $\Delta \mathrm{H}_{ij}$ & Amplitude difference of the blending-induced propagation per-baseline visibility error \\[2pt]
  $\Delta \Phi_{ij}$ & Phase difference of the blending-induced propagation per-baseline visibility error \\[2pt]
  $\bm{\mathsf{A}}_R, \bm{\mathsf{A}}_I$ & Array configuration matrix of the amplitude and phase equations, respectively \\[2pt]
  $\bm{\mathsf{A}}$ & Array configuration matrix $\bm{\mathsf{A}}$ in its general form \\[2pt]
  $\bm{\mathsf{A}}^\dagger$ & Pseudoinverse of the array configuration matrix $\bm{\mathsf{A}}$ \\[2pt]
  $\bm{b}$ & Gain factor difference, $R_{ij}$ matrix or $I_{ij}$ matrix, in its general form \\[2pt]
  $\bm{b}_R, \bm{b}_I$ & Amplitude difference $R_{ij}$ matrix and phase difference $I_{ij}$ matrix, respectively \\[2pt]
  $\bm{x}$ & Antenna gain matrix, amplitude or phase matrix, in its general form \\[2pt]
  $\bm{x}_R, \bm{x}_I$ & Antenna gain amplitude and phase matrix, respectively \\[2pt]
  $\bm{\tilde{x}}$ & Least-squares solution of the $\bm{x}$ matrix in its general form \\[2pt]
  $\bm{\tilde{x}}_R, \bm{\tilde{x}}_I$ & Least-squares solution of the gain amplitude and phase matrix, respectively \\[2pt]
  $\langle ... \rangle$ & Time average operation \\[2pt]
  \hline
 \end{tabular}
\end{table*}


\bsp	
\label{lastpage}
\end{document}